\documentclass[apj,twocolumn,numberedappendix]{openjournal}
\usepackage{natbib} 
\usepackage[dvipsnames]{xcolor}
\usepackage{amssymb}
\usepackage{amsmath}
\usepackage[caption=false]{subfig}
\usepackage{orcidlink}
\usepackage{hyperref}	

\newcommand{\simon}[1]{#1}

\hypersetup{colorlinks=true,linkcolor=red,citecolor=blue,filecolor=blue,urlcolor=blue}

\begin{document}
\title{The Impact of Non-Gaussian Line Spread Functions on Stellar Kinematic Recovery: Consequences for Dynamical Models\vspace{-15mm}}
\author{David A. Simon\orcidlink{0000-0001-5742-2982}$^{1,2*}$}
\thanks{$^*$E-mail: \href{mailto:davidsimon@westlake.edu.cn}{davidsimon@westlake.edu.cn}}

\affiliation{$^{1}$Sub-department of Astrophysics, University of Oxford, Keble Road, Oxford OX1 3RH, United Kingdom}
\affiliation{$^{2}$Department of Astronomy, Westlake University, Hangzhou 310030, Zhejiang Province, People’s Republic of China}

\begin{abstract}

\noindent The line spread function (LSF) of a spectrograph encodes the inherent broadening of a single spectral line. It is typically reported as a single number, the resolving power $R = \lambda/\Delta\lambda$ with $\Delta \lambda$ the FWHM of the LSF. In standard pipelines for extracting stellar kinematics the LSF is assumed to be a wavelength dependent Gaussian. However, detailed LSF measurements from real integral field spectrographs reveal a variety of shapes, some close to Gaussian, others with large wings or that appear boxy. I have studied the impact that these non-Gaussian LSF profiles have on the recovery of the stellar kinematics of a mock spectrum \simon{based on MUSE ($\sigma_{\rm inst} = 51 $ km s$^{-1}$)} and find that even in the high dispersion case of 300 km s$^{-1}$, there is up to a 7 percent \simon{bias} in the dispersion due to non-Gaussian LSF profiles. Additionally, higher order Gauss-Hermite moments $h_3$ and $h_4$ can be biased by up to $\pm$0.1. To resolve this bias, I developed a method to match the LSF of the template spectra to the LSF of a target spectrum when the LSF of either one or both is non-Gaussian and show that it can reduce bias in the dispersion to less than a percent down to the instrumental resolution. A Python implementation of this method has been made publicly available.
\end{abstract}

\section{Introduction}
\subsection{Background}
Standard approaches for extracting stellar kinematics from integral field data involve fitting a template spectrum to a galaxy or target spectrum by applying a velocity shift and broadening to the template spectrum (e.g. \cite{cappellari2004parametric,cappellari2017improving,cappellari2022full,wilkinson2017firefly,cidfernandes2005starlight,sanchez2016pypipe3d,lacerda2022pypipe3d}). However, kinematics are not the only source of broadening. Another key source is the spectral point spread function, also known as the line spread function (LSF). The LSF encodes the broadening of single spectral lines at each wavelength of the spectrum. In order to ensure that the kinematics measured with a spectral fitting code are accurate, it is necessary to first ensure that the template spectra and target spectrum have the same LSF.

The LSF is typically reported as a single number in terms of the resolving power of the spectrograph, defined as
\begin{equation}
	R = \frac{\lambda}{\Delta \lambda}
\end{equation}
where $\Delta \lambda$ is the full width at half maximum (FWHM) of the LSF. Most measurements of stellar kinematics are made assuming that the intrinsic shape of the LSF is Gaussian, and relate the FWHM to the standard deviation of a Gaussian using the relationship
\begin{equation}
	\Delta\lambda = 2 \sqrt{2 \ln 2}\sigma_\lambda \approx 2.355 \sigma_\lambda
\end{equation}
Under this assumption, the LSF of the template spectra is matched to the LSF of the target spectrum by taking a convolution of the form
\begin{equation}
	\text{LSF}_{\rm Template} * G = 	\text{LSF}_{\rm Target} 
\end{equation}
where $G$ is the function that matches the template LSF to the target LSF. In the case that both $\text{LSF}_{\rm Template}$ and $\text{LSF}_{\rm Target}$ are Gaussian, $G$ can be solved for exactly and is a Gaussian with a standard deviation given by
\begin{equation}\label{eq:gauss_conv}
	\sigma = \sqrt{\sigma_{\rm Target \ LSF}^2 - \sigma_{\rm Template \ LSF}^2}
\end{equation}
In general this standard deviation is wavelength dependent and is applied to the template spectrum with a wavelength dependent convolution code (see \cite{cappellari2022full}, Algorithm 1). This expression also implies that this matching convolution can only be performed when $\sigma_{\rm Target \ LSF} > \sigma_{\rm Template \ LSF}$. Otherwise the target spectrum must be broadened to match the template spectrum. 

\subsection{Origins of the LSF}
There are strong theoretical reasons to expect that the LSF may not be described as a perfect Gaussian. For slicer-based integral field spectroscopy (IFS), the LSF, to first order, can be described as a convolution of four separate terms: the slit, the diffraction grating dispersion, the instrument PSF (also called the camera PSF), and a pixel wide top hat function (this converts the output to something measured on a pixelated detector). If the slit is uniformly illuminated then the contribution from it is a top hat function with width equal to the slit width (typically 2-3 times the size of a spectral pixel). The case where the slit is not uniformly illuminated gives rise to the so-called classical slit effect and can result in significant uncertainties (e.g. section 2 of \cite{bacon19953d}). However, in the case of studies of galaxies and stellar kinematics the slit is likely to be close to uniformly illuminated so this is not likely to be a significant contributor to the LSF shape. Convolving the slit with the pixel top hat then gives a trapezoid. The diffraction grating dispersion is typically a fraction of a pixel wide, and thus small compared to the other terms\footnote{Though in some cases, such as for SINFONI, it can be significant \citep{george2017spiffi}.}. The convolution with the instrumental PSF is the last step in determining the (first order) shape of the LSF. For a Gaussian instrumental PSF, this will produce a boxy LSF. For an instrumental PSF with wide wings, this can produce an LSF with wings. Fiber based IFS and lenslet based IFS are similar to this, except the convolution with the slit is replaced by a convolution with the projected fiber core, or a convolution with the exit micro-pupil respectively. These effectively remove any uncertainty from the slit effect. Therefore, in each of the major varieties of IFS there is freedom in the final shape of the LSF depending on the specific characteristics of the instrument (see \cite{bacon2017optical} Sec. 8.2 for more details).

\subsection{Empirical LSFs}\label{sec:empirical_lsf}
In general, making detailed measurements of the LSF for real instruments is difficult. Integral field spectrographs are typically designed to just satisfy the Nyquist-Shannon sampling criterion \citep{nyquist1928certain,shannon1949communication}, making detailed LSF profiles difficult to measure in a single observation with the spectrograph. Several techniques have been devised to measure detailed LSF profiles. One approach is to take multiple observations of standard spectral lines while manually shifting the diffraction grating by a fraction of a pixel, thus super-sampling the LSF \citep{thatte2012sinfoni,kakkad2022lsf,kravchenko2022firsteris}. This provides significant control over how much over-sampling is performed, but its feasibility depends on the design of the instrument\footnote{\cite{kakkad2020fabry} proposes a method to perform hyperspectral sampling without making any internal modifications to the spectrograph using a Fabry-P{\'e}rot etalon.}. Another approach is to use slit or line curvature \citep{kelson2003optimal,kakkad2022lsf,law2021mangalsf}. Curvature refers to the fact that single spectral lines (e.g. from calibration lamps), will have different centroid locations within a pixel at different areas on a detector. This is effectively the same as manually shifting the grating though the shift is set by the instrument design and does not always uniformly sample the LSF \cite[Fig. 10]{kakkad2022lsf}. A final approach when neither of those options are available is to directly measure the width of \simon{unresolved or intrinsically very narrow} observed lines by fitting either a Gaussian or Gauss-Hermite series. This does not allow for detailed profiles of the LSF to be determined, but it does allow sufficient generality for non-Gaussian features to be detected.

The LSF for SINFONI \citep{eisenhauer2003sinfoni} has been measured by manually shifting the diffraction grating by a fraction of a pixel to hyper-sample single spectral lines \citep{thatte2012sinfoni,kakkad2022lsf}. They found that LSF profiles are generally strongly non-Gaussian, and have variations in their shape as a function of both wavelength and spatial location on the detector. They parameterise their LSF profiles as a Gauss-Hermite series \citep{vandermarel1993gausshermite,gerhard1993line} and find that terms up to $h_{12}$ must be included for an accurate fit. They investigate variations as a function of wavelength by plotting the different GH moments across the wavelength range and find the variations to be quite smooth \cite[Fig. 9]{kakkad2022lsf}.

In ERIS \citep{davies2023eris}, the successor instrument to SINFONI, the shape of the LSF has been measured in the same way as SINFONI and has significantly improved, with a shape that is close to Gaussian \cite[Fig. 6]{kravchenko2022firsteris}, though in some cases it can have an asymmetric tail \cite[Fig. 26]{dallilar2022erismanual}. 

The LSF of KCWI on Keck \citep{morrissey2018kcwi} has been measured by simultaneously fitting lines from an FeAr arc lamp. There it was found that the LSF is distinctly non-Gaussian, being better described as the convolution of a Gaussian with a top hat function \cite[Fig. A1]{liepold2023m87}. 

The LSF of JWST NIRSpec  \citep{boker2022nirspec} has been determined by two separate groups measuring the broadening of H and He lines in planetary nebulae. This is a natural choice for measuring the LSF as these lines have low intrinsic dispersion. In both cases the recovered LSFs are close to Gaussian though there are occasionally some asymmetries in the tails of the distribution \cite[Fig. 1]{isobe2023jwstlsf}. In the other case the authors note that fits to the LSF using a Voigt profile (approximately Gaussian with tails) perform better than a pure Gaussian \citep{shajib2025lsf}. 

For the fiber based IFS SAMI \citep{croom2012sami}, the non-Gaussian shape of the LSF was measured by fitting calibration lamp lines allowing for higher order Gauss-Hermite moments $h_3$ and $h_4$ \citep{vandesande2017higherorder}. They found that across their fits, $h_3$ and $h_4$ typically have a median less than 0.01, suggesting an LSF shape that is close to Gaussian. 

One of the most detailed studies of the LSF for a fiber based integral field spectrograph was made for MaNGA \citep{bundy2015manga}. There, they measured the LSF using curvature and found that the shape of the LSF was close to Gaussian, though a pure Gaussian slightly under-estimated the wings of the distribution \cite[Fig. 2]{law2021mangalsf}. They also found that the LSF depends smoothly on the wavelength, though different fiber ids sometimes exhibit sharp changes in the LSF width \cite[Fig. 3]{law2021mangalsf}. They also observed some time dependence of the LSF, possibly due to gravitational flexure in the fibers, varying camera optics, and variations in the detector focal plane. In addition to these short-term variations in the LSF, there are also long-term variations with different data releases \cite[Fig. 13]{law2021mangalsf}. One key test they perform is to validate the LSF profiles derived from calibration lamps with measurements of the LSF made by fitting sky lines. They find differences in the standard deviation of the Gaussian they fit between these two approaches of up to 10 percent. 


In addition to non-Gaussian LSF profiles playing a potential role in integral field observations, they are also important for stellar template libraries. Despite this, detailed LSF profiles for stellar template libraries are typically not measured. For example, in the X-Shooter stellar library \citep{chen2014xshooter}, the LSF is measured assuming a Gaussian LOSVD by fitting synthetic spectra to real X-Shooter observations of standard stars. Likewise for the MILES stellar library \citep{sanchez2006medium,falcon2011updated}, the spectral resolution was determined assuming a Gaussian fit to either calibration lamp spectra or by comparing high resolution spectra to a standard star assuming a Gaussian broadening. In all of these cases pure Gaussians are assumed without the option for non-Gaussian shapes. This creates the potential for a `double mismatch', where the LSF of a target spectrum is different from the LSF of the template spectra in a way that magnifies the uncertainty of the recovered kinematics. One way for this to happen would be if the target spectrum has an LSF with wings whereas the template spectra are more flat-topped. In this situation, the template spectrum would have to have a wide dispersion first to account for the boxiness of the original LSF, and then be broadened further to match the wings of the target spectrum. 

\subsection{Problem Statement and Approach}
That LSF profiles can be treated as a pure Gaussian is a fundamental assumption of nearly all spectral fitting approaches, both from the template side and the observed spectrum side. However, non-Gaussian shapes of the LSF appear to be a regular feature of integral field spectroscopy. The goal of this paper is to investigate the assumption of a pure Gaussian LSF profile and determine what systematic effects this induces.

This paper is structured as follows: In \autoref{sec:mock_data}, I produce a realistic template and target spectrum with wavelength-dependent non-Gaussian LSF profiles. I describe the details of how these mock observations are fitted with \textsc{pPXF} \simon{\citep{cappellari2004parametric,cappellari2017improving,cappellari2022full}} in \autoref{sec:kin_extra} and show the results for this in \autoref{sec:recov_kin}. I discuss different strategies for mitigating this effect in \autoref{sec:mitig_strat} and describe a detailed approach in \autoref{sec:solve_conv}. Lastly, I discuss these results in the context of other observations in \autoref{sec:discussion} and close in \autoref{sec:conclusion} with the importance of making detailed LSF measurements and integrating them into existing spectral fitting pipelines.

\section{Mock Data}\label{sec:mock_data}
\subsection{Template and Target Spectra}

I base the template and target spectrum on the single stellar population models of \cite{maraston2011stellar}, based on the MARCS synthetic library \citep{gustafsson2008marcs}. This library has very high spectral resolution (R = 20000) spanning a wide wavelength range from nearly 1000 \AA \ to 25000 \AA. I use a single SSP with age equal to 10 Gyr, metallicity [Fe/H] = 0.0, and a Salpeter IMF \citep{salpeter1955imf}. A typical real observation will be subject to multiple sources of uncertainty. These include Poisson noise from the observation, template mismatch, etc. In order to best isolate the impact from non-Gaussian and wavelength-dependent LSF profiles, I only consider an idealised case where the galaxy spectrum and template spectrum are constructed from the same underlying template without added noise or biases. This guarantees that the bias in the recovered kinematics derives purely from the LSF.  

\subsection{Instrumental Characteristics}
\begin{figure}
	\centering
	\subfloat{\includegraphics[width = 1\columnwidth]{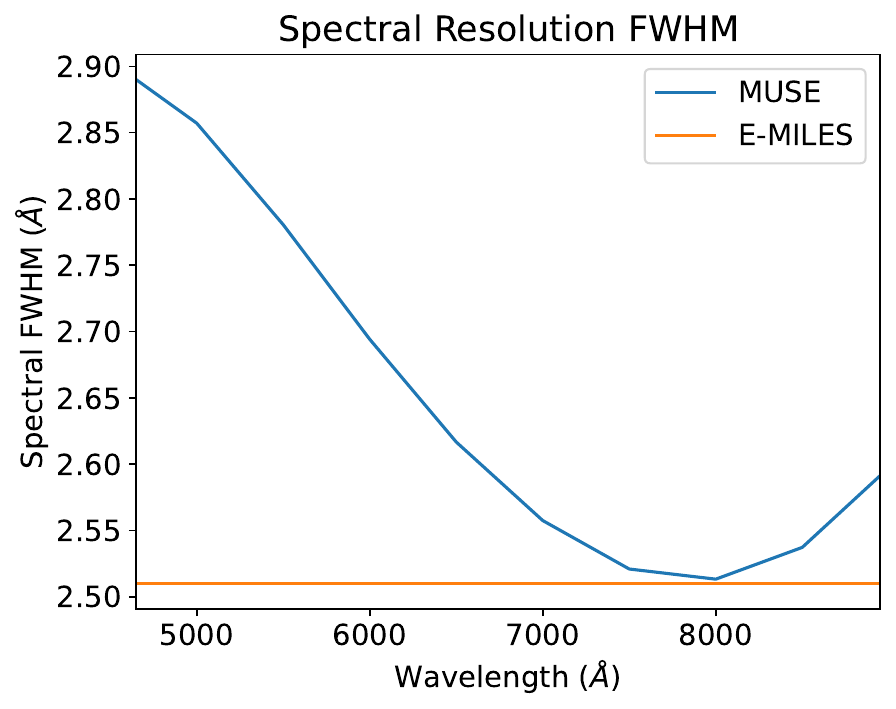}}\\
	\subfloat{\includegraphics[width = 1\columnwidth]{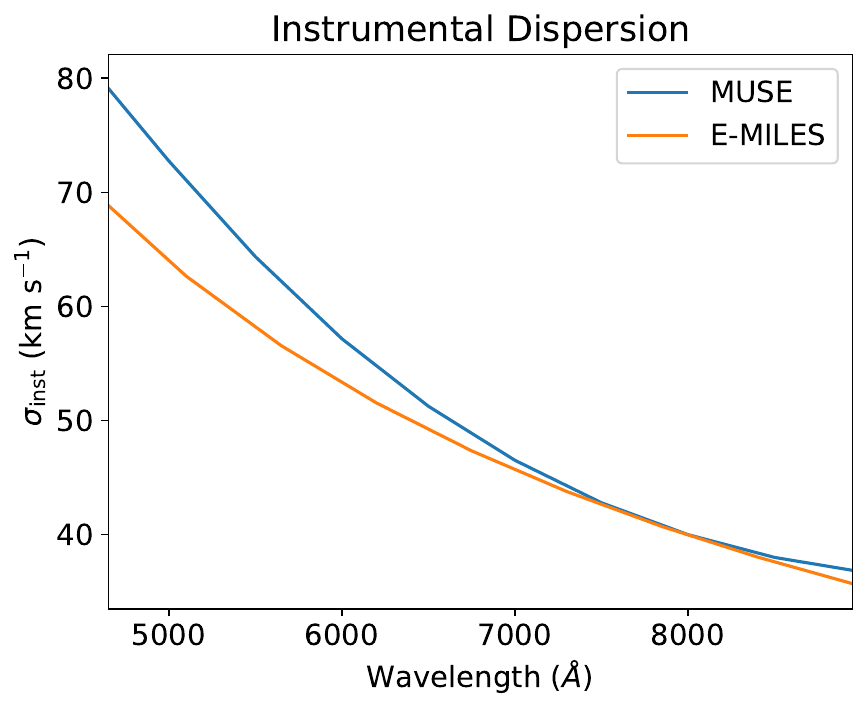}}
	\caption[]{Spectral resolution in FWHM and $\sigma_{\rm inst}$ for MUSE and E-MILES as a function of wavelength. Taken from figure 18 of the MUSE user manual \citep{richard_2026MUSE_UserManual} and table 1 of \cite{vazdekis2016emiles}. }
	\label{fig:MUSE_spec_res}
\end{figure}

The mock target spectrum is modeled on the instrumental characteristics of MUSE \citep{bacon2010muse} and the template spectrum on the E-MILES stellar library \citep{vazdekis2016emiles}. MUSE has a wavelength range of 4800 \AA \ to 9300 \AA \ and a spectral resolution in FWHM ranging roughly between 2.51 \AA \ and 2.89 \AA. The E-MILES library is constructed by combining multiple templates together and thus has a large wavelength range from 1680 \AA \ to 50000 \AA \ and a non-continuous spectral resolution. However, within a range close to the MUSE data (3541.4 \AA \ to 8950.4 \AA), the spectral resolution is a constant 2.51 \AA \ \cite[Table 1]{vazdekis2016emiles}. Thus to simplify the analysis I clip the wavelength range of the target spectrum to the range 4800 \AA \ to 8950 \AA \ (in \autoref{sec:uncertain_conv} I show that this does not meaningfully impact the recovery of the kinematics). Plots of the spectral resolution are shown in \autoref{fig:MUSE_spec_res}.

The MARCS library is logarithmically spaced in wavelength. In order to simplify the convolution with the LSF I interpolate the spectrum onto a linearly spaced wavelength region with spectral pixel length equal to 1.25 \AA \ (the MUSE value). \simon{The reason for interpolating is that the LSF is measured as a linear function of wavelength. A simple Gaussian LSF then takes on a complicated shape when logarithmically spaced making the LSF convolution less accurate.} The MARCS spectral pixel size on this wavelength range spans from 0.24 \AA \ at 4800 \AA \ and 0.45 \AA \ at 8950 \AA \ so the interpolation is well sampled. However, it is worth emphasising that even degrading the resolution of the spectrum would not impact the results as this linearly sampled spectrum is the `true' spectrum that is fitted. 

\subsection{Realistic LSF Profiles}\label{sec:real_lsf}
\begin{figure}
	\centering
	\subfloat{\includegraphics[width = 1\columnwidth]{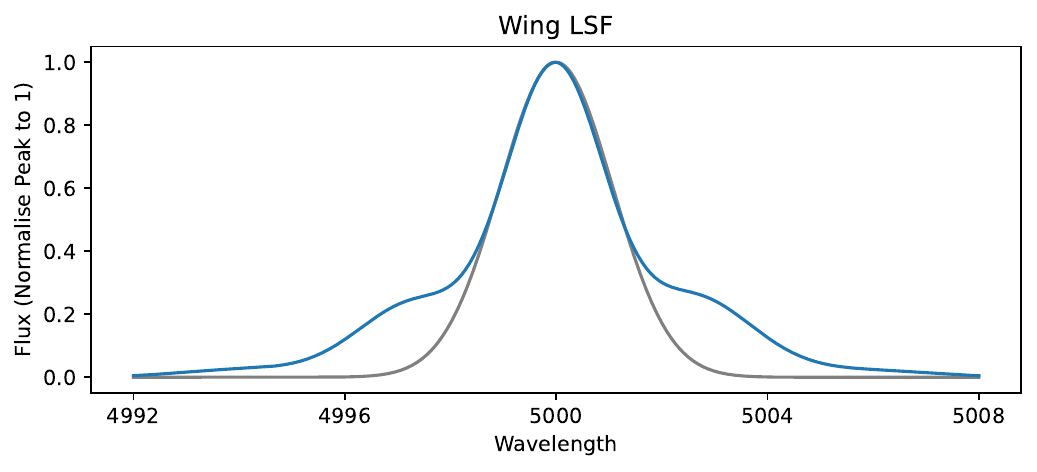}}\\
	\subfloat{\includegraphics[width = 1\columnwidth]{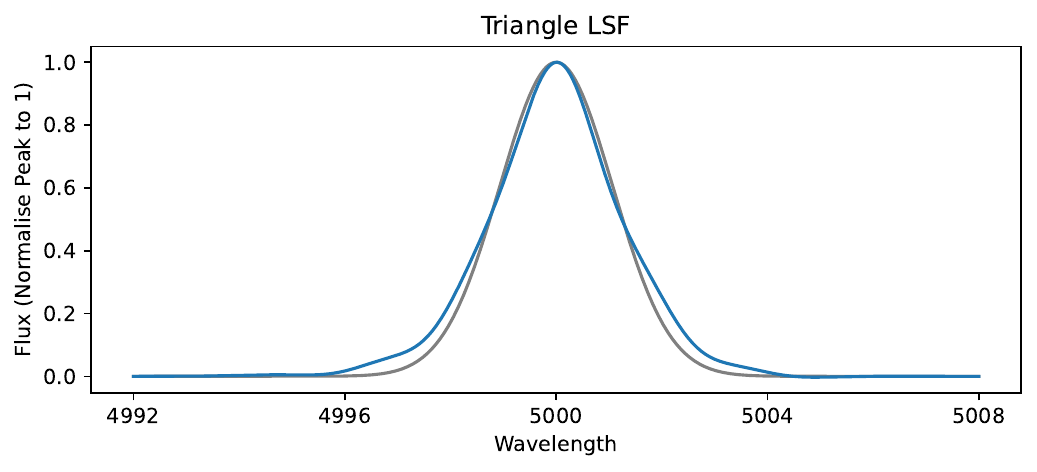}}\\
	\subfloat{\includegraphics[width = 1\columnwidth]{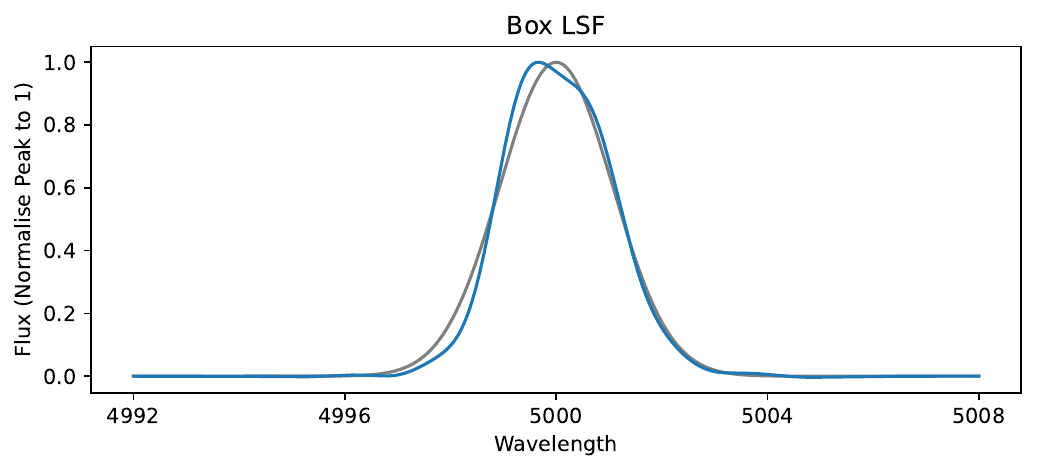}}
	\caption[]{Sample of three LSF profiles taken from the SINFONI handbook \cite[Fig. 22-25]{hau2017sinfoni}. Each LSF is centred at 5000 \AA \ and given a FWHM of 2.51 \AA. The curve in grey is the profile of a Gaussian with the same FWHM as the sample LSF profile. It is clear that even at the same FWHM there is significant variation in the shape of the profiles.}
	\label{fig:lsf_profiles}
\end{figure}

The next step to create a realistic spectrum is to apply the LSF. As seen in \autoref{sec:empirical_lsf}, there is a variety of LSF shapes realised at different integral field spectrographs. An extensive array of these different LSF shapes is shown in the SINFONI handbook \cite[Fig. 22-25]{hau2017sinfoni}. These can be broadly classified into three categories based on their morphology. There are the `winged' profiles, the `boxy' profiles, and the `triangular' profiles (in this work these are often shortened to W, B, and T for brevity). I extract a representative profile for each of these morphologies using the program \textsc{WebPlotDigitizer} \citep{WebPlotDigitizer} and have plotted the results with a Gaussian of the same FWHM in \autoref{fig:lsf_profiles}. For reasons described in \autoref{sec:lsf_convolution}, it is convenient to parameterise these profiles using a Gauss-Hermite (GH) series \citep{vandermarel1993gausshermite,gerhard1993line}. For each of the three profiles in \autoref{fig:lsf_profiles}, I fit a GH series of the form
\begin{equation}\label{eq:LSF_lambda}
	\text{LSF}(\lambda) = \frac{e^{\frac{-\Lambda^2}{2}}}{\sigma_\lambda \sqrt{2 \pi}}\left(1 + \sum_{n=3}^{12}h_nH_n(\Lambda)\right)
\end{equation}
where $\Lambda = (\lambda - \lambda_0)/\sigma_\lambda$. I truncate the series at $h_{12}$. This is the same value found in \cite{kakkad2022lsf} to provide an accurate fit to SINFONI LSF profiles. In addition to these three profiles, I also perform all of the tests with a Gaussian LSF.

\subsection{LSF Convolution}\label{sec:lsf_convolution}
Special care must be given to the convolution in order to ensure the accuracy of the convolved spectrum. A straightforward implementation of the convolution with direct summation will have low accuracy when the convolution kernel is smaller than the spectral pixel size. The solution to this is to first note that a convolution between two functions, $f$ and $g$, can be rewritten using the Fourier convolution theorem as
\begin{equation}
	f*g = \mathcal{F}^{-1}[\mathcal{F}[f]\mathcal{F}[g]]
\end{equation}
The problem of performing the convolution then reduces to evaluating these Fourier transforms. The benefit of this approach is that, in the standard direct summation case, sub-sampled kernels start to appear as box functions due to the size of the spectral pixel. This introduces errors to the convolution as the kernel size approaches the size of a spectral pixel. However, in the Fourier approach, for special choices of $f$ or $g$, the Fourier transform can be evaluated analytically, thus allowing accurate convolutions to be performed even when the kernel size is smaller than the pixel size. For functions without an obvious analytic Fourier transform (such as template or target spectra), these can still be evaluated using the fast Fourier transform \citep{cooley1965algorithm}.

The question then is how to choose a function or set of basis functions that allow a straightforward analytic Fourier transform to be calculated. It has previously been pointed out that Gauss-Hermite polynomials are eigenfunctions of the Fourier transform \cite[Eq. A7]{vandermarel1993gausshermite}. This is a very surprising and beautiful result, but it has a straightforward explanation. The Hamiltonian for the quantum harmonic oscillator is given by the following differential operator
\begin{equation}
	\hat{H} = -\partial_x^2 + x^2
\end{equation}
The Fourier transform (in the modern physics convention) has the property that $\partial_x \to ik$ and $x \to -i\partial_k$. This implies that
\begin{equation}
	\mathcal{F}[\hat{H}] =  - \partial_k^2 + k^2
\end{equation}
Therefore, the Fourier transform commutes with the quantum harmonic oscillator Hamiltonian, and thus they share a common eigenbasis. As the eigenbasis for the quantum harmonic oscillator is the Hermite polynomials \cite[Eq. 26]{Schrödinger1926QaE}, the Fourier transform also has Hermite polynomials as its eigenfunctions\footnote{\simon{For another approach with an analytic Fourier transform, see \cite{sanders2020neargaussian}.}}.

A numerical implementation of the analytic convolution of Gauss-Hermite functions with an arbitrary function is provided in the \textsc{convolve\_gauss\_hermite} function of the \textsc{pPXF} package, and is described in \cite{cappellari2017improving} where it is used in \textsc{pPXF}\footnote{Publicly available from \url{https://pypi.org/project/ppxf/}} to evaluate the convolution of template spectra with a line of sight velocity distribution parameterised using the GH series.

With this, it is now possible to accurately convolve a function parameterised as a GH series with a template. This handles the non-Gaussian aspect of the LSF. However, LSF profiles are not just non-Gaussian but also wavelength dependent. A full description of the approach for solving the wavelength dependent version of this problem is given in algorithm 1 of \cite{cappellari2022full}. The key idea is that the spectrum is stretched with interpolation such that the broadening is constant over the stretched spectrum. Then the analytic GH convolution can be applied as in the previous part. \textsc{pPXF} implements this for the pure Gaussian case in the function \textsc{varsmooth}, but I have extended this to the case with the full Gauss-Hermite series by combining this function with parts of the \textsc{convolve\_gauss\_hermite} function. I have tested that this code behaves as expected by generating a mock spectrum that consists of a series of equally spaced delta functions and confirmed that the FWHM of the convolved spectrum matches the target FWHM. 

One drawback of this approach is that it introduces two interpolations, one onto a grid where $\sigma$ is equally spaced and one back to the original spacing of the spectrum. This necessarily introduces some uncertainty into the final convolved spectrum. There are two ways to minimise the error from this. The first is to oversample the interpolated spectrum. I find diminishing returns beyond an oversampling by a factor of 10. Second, different interpolation methods can be chosen. I find that the most accurate results can be achieved with a cubic spline interpolation, rather than a linear interpolation. I implement this using \textsc{scipy}'s \textsc{CubicSpline} function. A detailed discussion of this and the errors introduced by interpolation is given in \autoref{sec:uncertain_conv}.

The final step for the template spectrum is to perform a convolution to match the spectral resolution of the template to the target spectrum. This is done in the standard way by assuming they can be matched using a Gaussian convolution. This is performed using the \textsc{varsmooth} function with a broadening given by 
\begin{equation}
	\sigma(\lambda) = \sqrt{\sigma^2_{\rm target}(\lambda) - \sigma^2_{\rm template}(\lambda)}
\end{equation}
where $\sigma_{\rm target}$ and $\sigma_{\rm template}$ are related to the FWHM with $\sigma = 2\sqrt{2\ln 2}\times \text{FWHM}$.

To summarise, I assume a wavelength dependent LSF where the shape is given by one of the representative LSF shapes from \autoref{fig:lsf_profiles}, and the wavelength dependence is given by the spectral resolution profiles of \autoref{fig:MUSE_spec_res}. The wavelength dependence is applied by uniformly scaling the LSF shape to match the FWHM of the given spectral resolution profile. One limitation of this approach is that it requires that the shape of the LSF (up to a stretching) is constant over the spectral range. The SINFONI detailed LSF profiles show an approximately constant LSF shape across the wavelength range (winged LSFs retain their wing structure at all tested wavelength ranges, boxy LSFs retain their boxy structure at all tested wavelength ranges, etc.) \cite[Fig. 22-25]{hau2017sinfoni}. For this reason, this is likely to be an adequate assumption for the sake of these tests.

\subsection{Kinematic Broadening}
Once the LSF has been applied to the linearly spaced spectrum, the next step for the target spectrum is to apply the effect of kinematic broadening. This is done by first log rebinning the spectrum using the \textsc{log\_rebin} function of \textsc{pPXF}. I rebin to a velocity scale of 40.3 km s$^{-1}$. The spectrum is then convolved with a chosen LOSVD using the \textsc{convolve\_gauss\_hermite} function. Since convolutions are poorly behaved at the edges of the spectrum, I perform both the LSF convolution and the LOSVD convolution on an extended wavelength range of 4500 \AA \ to 9500 \AA \ and then clip to the desired wavelength range at the very end.

\subsection{Summary}
I have described the process for generating a mock target spectrum and mock template spectrum to test the impact of non-Gaussian LSF profiles on the recovery of the LOSVD. I summarise the procedure below:
\begin{itemize}
	\item Generate single MARCS SSP with age = 10 Gyr and [Fe/H] = 0.0.
	\item Linearly interpolate MARCS SSP on an extended wavelength range of 4500 \AA \ to 9500 \AA \ to a spectral pixel size of 1.25 \AA. 
	\item Convolve wavelength dependent LSF$_{\rm target}$ with linearly spaced MARCS spectrum. This is the target spectrum that is to be fit with \textsc{pPXF}. 
	\item Convolve wavelength dependent LSF$_{\rm template}$ with linearly spaced MARCS spectrum. Then convolve with a pure Gaussian to match the FWHM of the target spectrum. This is the template spectrum that is to be used to fit the target spectrum using \textsc{pPXF}. 
	\item Logarithmically rebin both spectra in wavelength.
	\item For the target spectrum to be fit with \textsc{pPXF}, convolve with a target LOSVD of choice and cut to the MUSE wavelength range.
\end{itemize}
With the target spectrum and template spectrum now generated, the next step is to perform spectral fits with \textsc{pPXF}.

\section{Kinematic Extraction}\label{sec:kin_extra}
The template spectrum is fitted to the target spectrum using the code \textsc{pPXF} \citep{cappellari2004parametric,cappellari2017improving,cappellari2022full}. For a single template spectrum, \textsc{pPXF} describes the target spectrum as
\begin{equation}
	T_{\rm Model} = T_{\rm Template}* \mathcal{L}(cx)
\end{equation}
where $x = \ln\lambda$ and $c$ is the speed of light. The line of sight velocity distribution (LOSVD) $\mathcal{L}(cx)$ is expressed in terms of a Gauss-Hermite (GH) \citep{vandermarel1993gausshermite,gerhard1993line} series of the form
\begin{equation}
	\mathcal{L}(v) = \frac{e^{-y^2/2}}{\sqrt{2\pi}\sigma}\left[1+\sum_{m=3}^{M}h_mH_m(y)\right]
\end{equation}
where $y = (v-V)/\sigma$, $H_m(y)$ are Hermite polynomials, and $h_m$ are constant coefficients of the GH series. One key choice is how many GH moments $M$ to include. I perform two runs, one where I fit a pure Gaussian, and one where I fit up to $M=4$. These are standard choices made when studying real galaxies. 

The other key parameter choice for the \textsc{pPXF} fits is the penalisation, which, in the code, is called the bias. The bias should be chosen so that kinematics can be recovered without offset across the velocity dispersion range of interest. Below this limit, the recovered GH moments should tend to zero. Since the goal here is to recover the kinematics across the full dispersion range, I systematically varied the bias until stellar velocity dispersions larger than the instrumental dispersion return unbiased GH moments, while input velocity dispersions smaller than the instrumental dispersion recover GH moments that are biased towards zero (see \cite{cappellari2004parametric} for details). I find that bias = 0.3 satisfies this criterion. I also tried varying the penalisation by an order of magnitude and found that it had a negligible impact on the recovered GH moments above the instrumental resolution. Beneath the instrumental resolution the Nyquist-Shannon theorem is no longer satisfied and the fits to the GH moments are ill-determined. I therefore proceed with the bias equal to 0.3. 

\textsc{pPXF} contains optional parameters to include additive or multiplicative Legendre polynomials. These are often useful when studying real galaxies to account for template mismatch or non-stellar emissions. However, for these idealised tests they are not necessary and therefore are not included in the fits.

\section{Recovered Kinematics}\label{sec:recov_kin}
\subsection{Pure Gaussian LOSVD}
\begin{figure}
	\centering
	\subfloat{\includegraphics[width = 1\columnwidth]{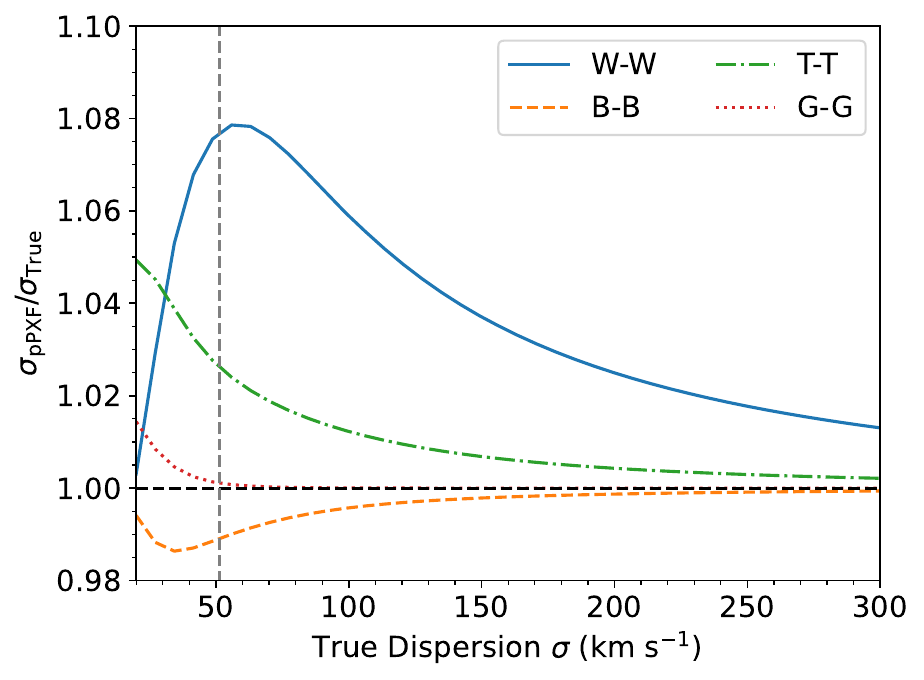}}
	\caption[]{Recovered \textsc{pPXF} $\sigma$ with $M=2$ for an underlying LOSVD that is a pure Gaussian. The notation W-W means the target spectrum is convolved with the winged LSF profile and the template spectrum is convolved with the winged LSF profile. The MUSE instrumental dispersion is the dashed gray line at $\sigma = 51.4$ km s$^{-1}$. The dispersion is always recovered to within 8 percent accuracy.}
	\label{fig:standard_dispersion_ww}
\end{figure}

First, I consider the case where the target spectrum has a pure Gaussian LOSVD with velocity 0. I choose the LSF profile for the target spectrum and template spectrum to be the same shape (but with a FWHM given by \autoref{fig:MUSE_spec_res}). Therefore the only difference between these two spectra when it comes to their spectral resolution is that the template spectrum is additionally convolved with a Gaussian as a function of wavelength to match the target spectrum FWHM. \simon{Unlike the pure Gaussian case, a Gaussian convolved with (for example) a winged profile does not produce a copy of the winged profile scaled to a larger FWHM, but a complicated new profile. This introduces some mismatch between the template and target LSF}. I vary the true stellar dispersion of the target spectrum and plot the recovered \textsc{pPXF} dispersion in \autoref{fig:standard_dispersion_ww} where I fit a pure Gaussian as the LOSVD ($M = 2$). For large values of the true stellar dispersion the recovery is accurate to 2 percent. Close to the instrumental dispersion there is more variation, with the maximum \simon{bias} of 8 percent. 

One case to pay special attention to is the G-G case, where a pure Gaussian is assumed for both the target and template spectrum. In this case, there is still some error in the recovery beneath the instrumental resolution. This is due to the uncertainty introduced by the wavelength dependent convolution that matches the instrumental dispersion of the template to the target spectrum. This is discussed in more detail in \autoref{sec:uncertain_conv}.

\begin{figure}
	\centering
	\subfloat{\includegraphics[width = 1\columnwidth]{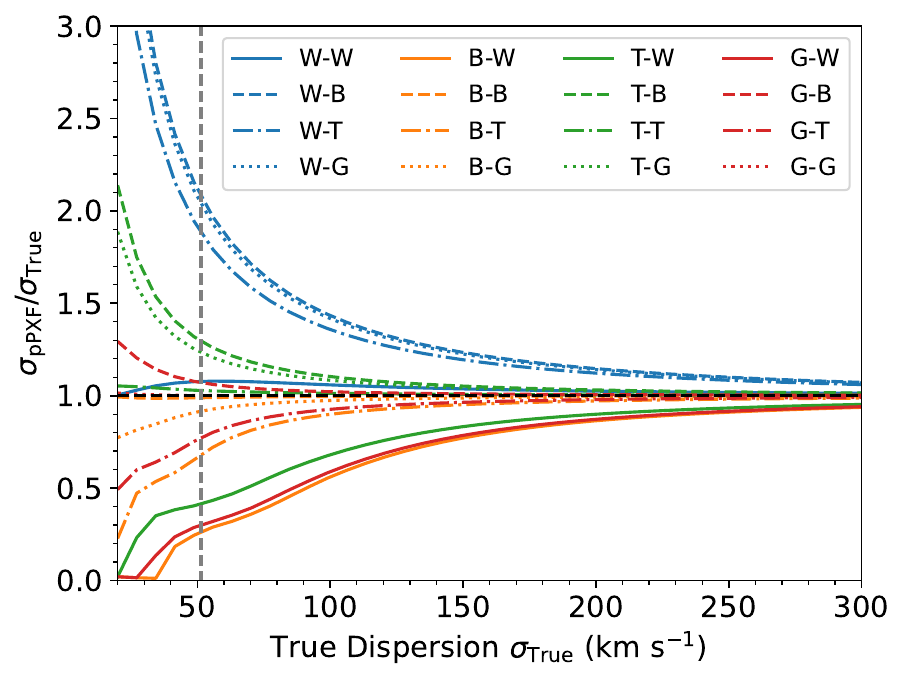}}
	\caption[]{\simon{Zoomed out version of \autoref{fig:standard_dispersion_ww} with more LSF combinations showing the} recovered \textsc{pPXF} $\sigma$ with $M=2$ for an underlying LOSVD that is a pure Gaussian. The notation W-W means the target spectrum is convolved with the winged LSF profile and the template spectrum is also convolved with the winged LSF profile. The MUSE instrumental dispersion is the dashed gray line at $\sigma = 51.4$ km s$^{-1}$. As expected, the most deviant recovery comes from the W-B and B-W cases since these maximise the `double mismatch' between LSF types.}
	\label{fig:standard_dispersion}
\end{figure}

Next, I show the results for this where I allow the template and target LSF to have any combination of different profiles. The result of this is shown in \autoref{fig:standard_dispersion}. The results here show much more variation than in \autoref{fig:standard_dispersion_ww}. The reason for this is that the previous case had target LSFs that were approximately equal to the template LSFs up to a broadening. However, in this new situation, the biases constructively add up to produce significant errors in the extracted stellar dispersion.

The range of behaviors observed by the recovered dispersion depends closely on the assumed LSF profiles. Fits where the target spectrum has a winged LSF naturally recover a much larger stellar dispersion as the template must expand larger to account for the large wings. Likewise, when the template is convolved with a winged LSF the dispersion is much smaller. For these fits, even at $\sigma = 300$ km s$^{-1}$ there is a \simon{bias} in the recovered dispersion of up to 7 percent. If the winged profiles are ignored then there is a maximum error at the instrumental dispersion of about 30 percent which decreases to 1.4 percent at 300 km s$^{-1}$.

\begin{figure}
	\centering
	\subfloat{\includegraphics[width = 1\columnwidth]{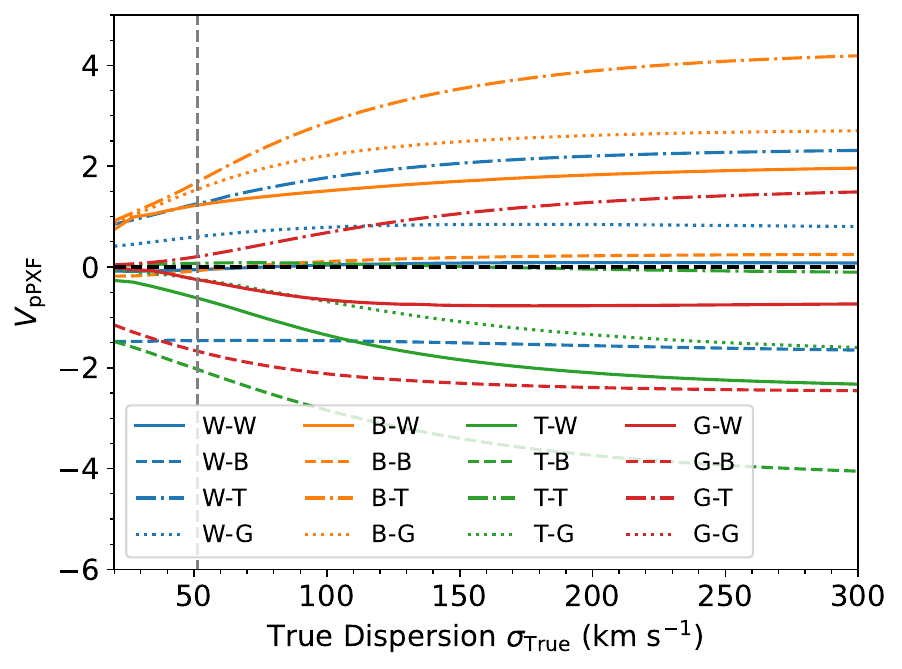}}
	\caption[]{Recovered \textsc{pPXF} $V$ for an underlying LOSVD that is a pure Gaussian. The notation W-W means the target spectrum is convolved with the winged LSF profile and the template spectrum is also convolved with the winged LSF profile. The MUSE instrumental dispersion is the dashed gray line at $\sigma = 51.4$ km s$^{-1}$. All spectra recover the velocity to within $\pm$4 km s$^{-1}$. This decreases with the dispersion down to $\pm$2 km s$^{-1}$.}
	\label{fig:standard_velocity}
\end{figure}

I also show the recovery of the velocity for different combinations of the LSF profile in \autoref{fig:standard_velocity}. There it is not the winged LSF that has the largest \simon{bias} but the boxy LSF. This is due to the boxy LSF having some clear asymmetries. Regardless, across the full range of velocity dispersion the \simon{bias} in the velocity only goes up to 4 km s$^{-1}$. Contrary to the previous case, the accuracy of the velocity measurements decrease with the true dispersion. The reason for this is that, while the convolution of the LOSVD with the LSF dilutes the asymmetric features of the LSF, the contribution of the asymmetric features to the velocity scale with the dispersion of the LOSVD. This results in a two scale behavior where the velocity initially increases before leveling off and eventually decreasing. A detailed explanation of this with an analytic toy model is given in \autoref{sec:inc_veloc}.

\begin{figure}
	\centering
	\subfloat{\includegraphics[width = 1\columnwidth]{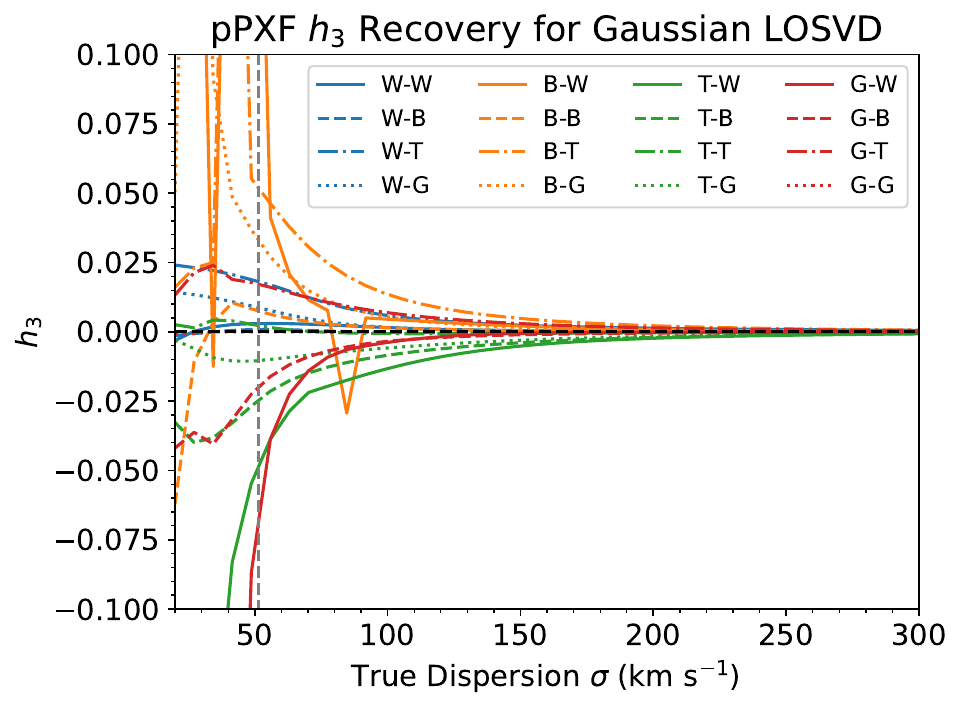}}
	\caption[]{Recovered \textsc{pPXF} $h_3$ for an underlying LOSVD that is a pure Gaussian. The notation W-W means the target spectrum is convolved with the winged LSF profile and the template spectrum is also convolved with the winged LSF profile. The MUSE instrumental dispersion is the dashed gray line at $\sigma = 51.4$ km s$^{-1}$. The true value is only recovered (within $\pm0.02$) for $h_3$ when $\sigma$ is 100 km/s, twice the instrumental dispersion.}
	\label{fig:h3_recov_gauss}
\end{figure}

\begin{figure}
	\centering
	\subfloat{\includegraphics[width = 1\columnwidth]{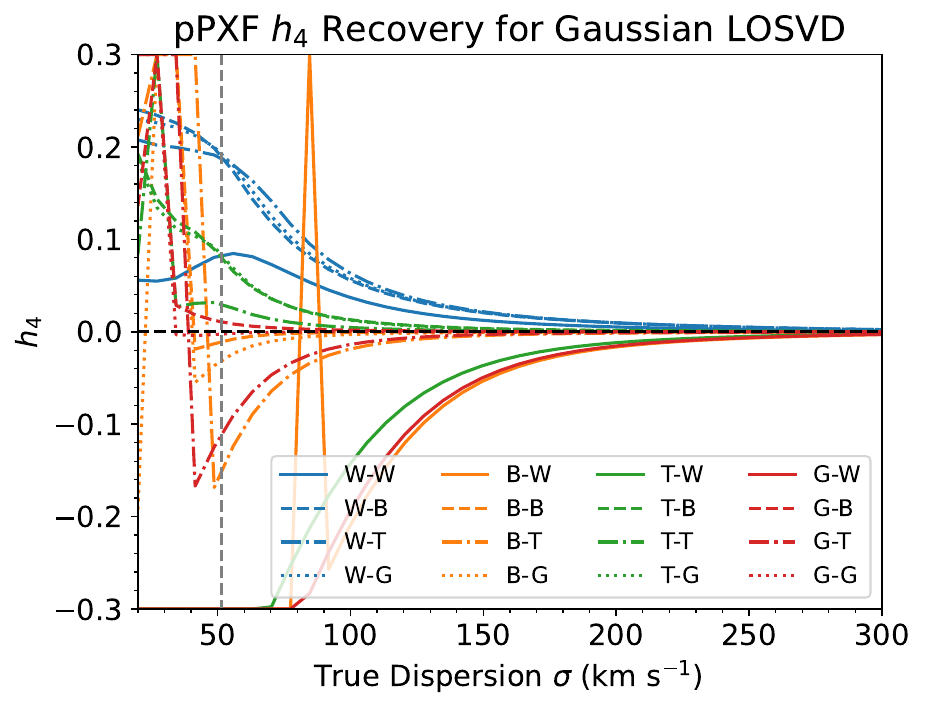}}
	\caption[]{Recovered \textsc{pPXF} $h_4$ for an underlying LOSVD that is a pure Gaussian. The notation W-W means the target spectrum is convolved with the winged LSF profile and the template spectrum is also convolved with the winged LSF profile. The MUSE instrumental dispersion is the dashed gray line at $\sigma = 51.4$ km s$^{-1}$. We see that the true value is only recovered (within $\pm0.02$) for $h_4$ when $\sigma$ is 200 km/s, four times the instrumental dispersion.}
	\label{fig:h4_recov_gauss}
\end{figure}

Next, I perform \textsc{pPXF} fits on the target spectrum with a pure Gaussian LOSVD but with $M$ = 4. I plot the recovery of $h_3$ and $h_4$ in \autoref{fig:h3_recov_gauss} and \autoref{fig:h4_recov_gauss}, respectively. $h_3$ is typically recovered much better than $h_4$ as the LSF shapes mostly change the shape of the wings and have small asymmetries. One important feature is that in some cases the recovered values of $h_3$ and $h_4$ start significantly diverging before reaching the instrumental dispersion. The reliable region ends approximately around twice the instrumental dispersion. It should be noted that this divergent behavior of $h_3$ and $h_4$ occurs even though the \textsc{pPXF} bias is included. The reason for this is that the bias is not meant to account for systematic uncertainties in the GH moments, but rather is meant to ensure the GH moments aren't completely unconstrained in the undersampled limit of an unbiased spectrum. As the LSF does bias the results, this failed recovery of $h_3$ and $h_4$ is not unexpected.

\begin{figure}
	\centering
	\subfloat{\includegraphics[width = 1\columnwidth]{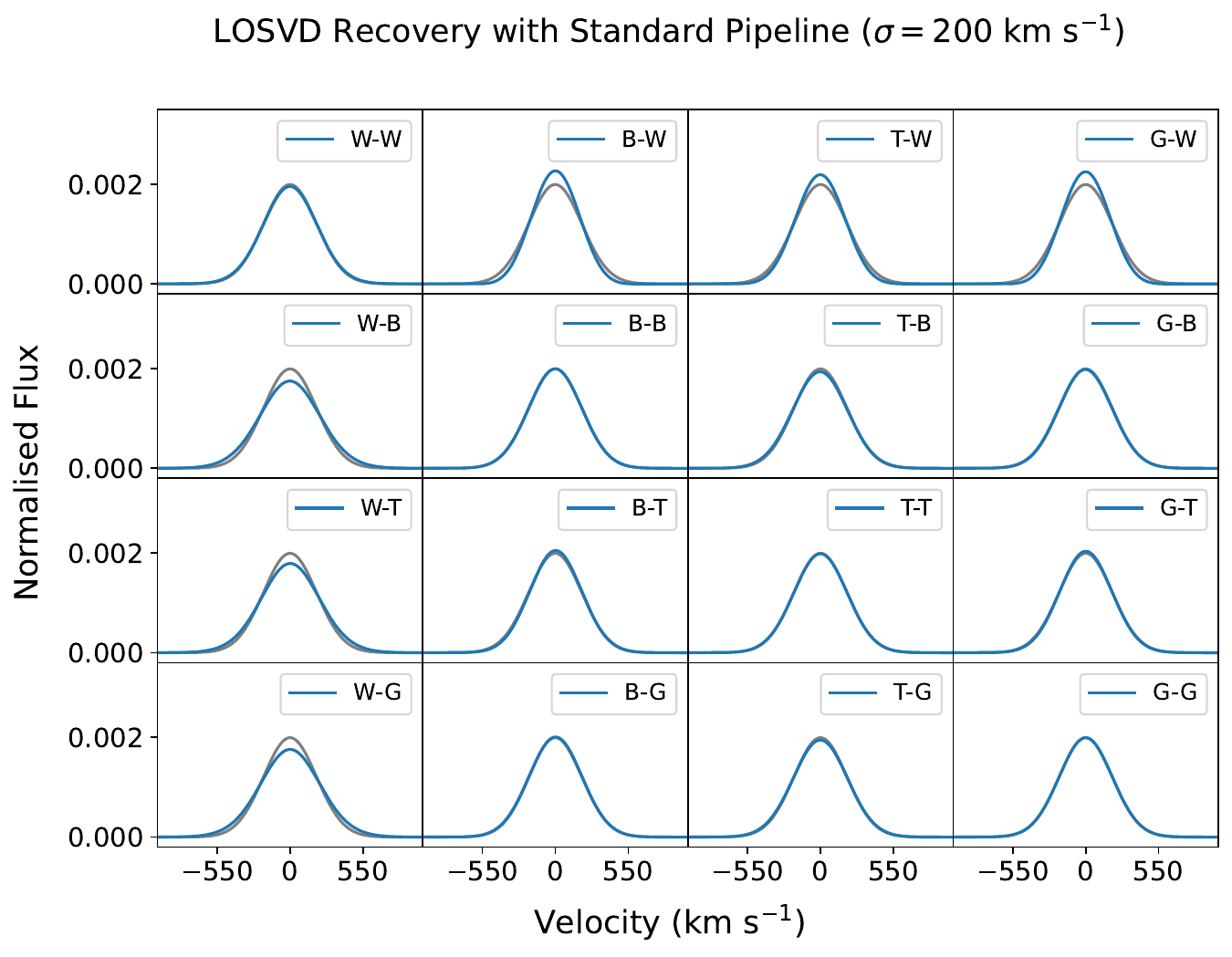}}
	\caption[]{Recovered \textsc{pPXF} LOSVD shapes with $M=4$ for an underlying stellar LOSVD that is a pure Gaussian with $\sigma = 200$ km s$^{-1}$. The true LOSVD is shown in gray. The notation W-W means the target spectrum is convolved with the winged LSF profile and the template spectrum is also convolved with the winged LSF profile. Plots along the diagonal naturally give the best recovery while off diagonal plots have stronger errors.}
	\label{fig:SS_losvd_recovery1}
\end{figure}

\begin{figure}
	\centering
	\subfloat{\includegraphics[width = 1\columnwidth]{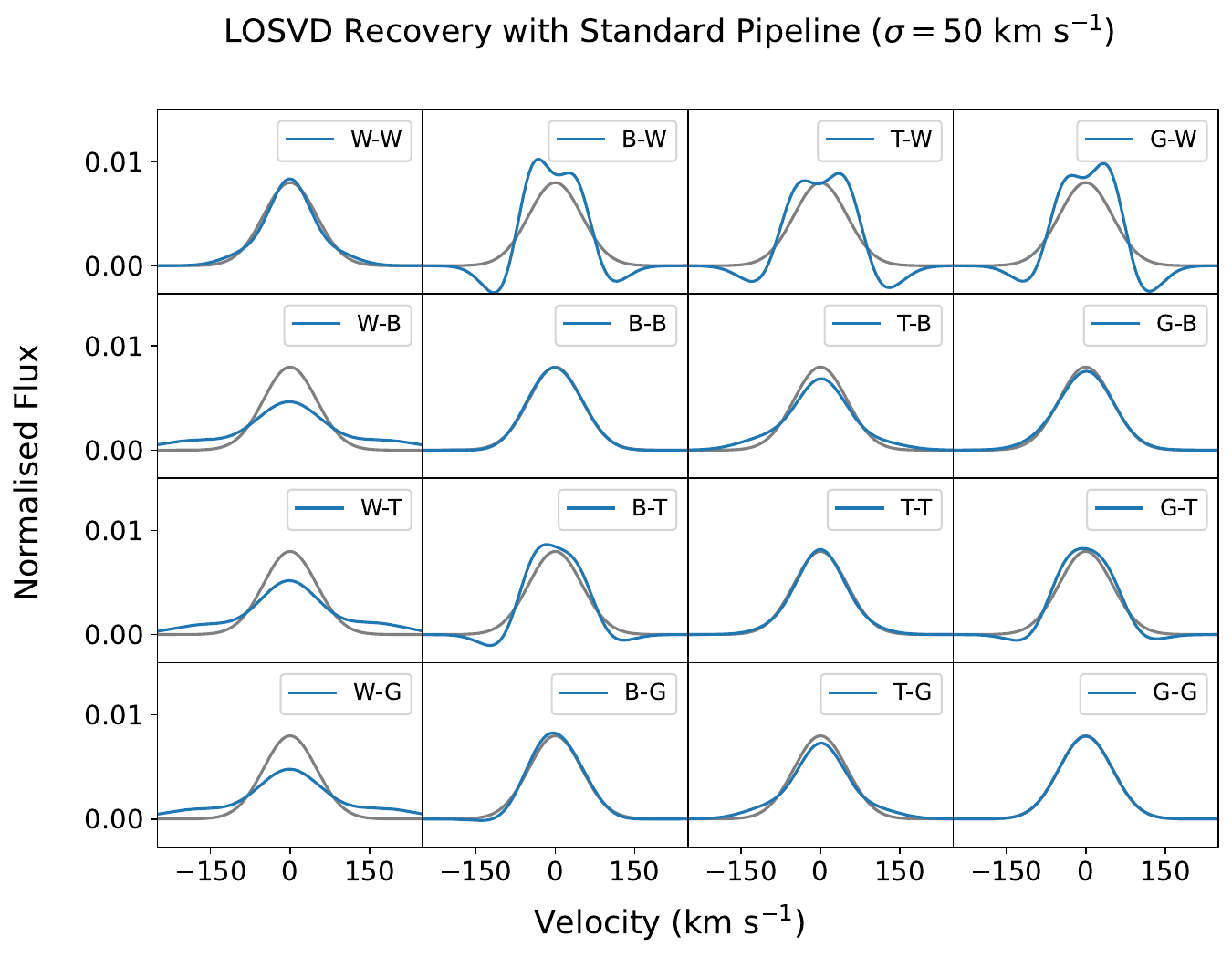}}
	\caption[]{Recovered \textsc{pPXF} LOSVD shapes with $M=4$ for an underlying stellar LOSVD that is a pure Gaussian with $\sigma = 50$ km s$^{-1}$. The true LOSVD is shown in gray. The notation W-W means the target spectrum is convolved with the winged LSF profile and the template spectrum is also convolved with the winged LSF profile. Plots along the diagonal naturally give the best recovery while off diagonal plots have stronger errors. The most deviant profiles are the ones involving the winged spectrum since this most strongly impacts the recovery of the wings of the LOSVD.}
	\label{fig:SS_losvd_recovery2}
\end{figure}

To get a better idea of how the shape of the LOSVD responds to different shapes of the LSF, I show a plot of the bestfit LOSVD profiles where I fit $M=4$ for $\sigma = 50$ km s$^{-1}$ and $\sigma = 200$ km s$^{-1}$ in \autoref{fig:SS_losvd_recovery1} and \autoref{fig:SS_losvd_recovery2}, respectively. There, it is immediately clear that even in the large dispersion case with $\sigma = 200$ km s$^{-1}$ the recovered shape of the LOSVD depends on the detailed LSF profile. Close to the instrumental dispersion, the profiles are very strongly impacted depending on the combination of template and target LSF profile. Combinations where the template and target have the same shape are always the least affected, whereas profiles including the winged LSF are the most affected. The behavior is as expected: winged target LSFs recover wider LOSVDs and winged template LSFs recover sharper, more boxy LOSVDs with negative wings.

\subsection{LOSVD with $h_3$ and $h_4$}
\begin{figure}
	\centering
	\subfloat{\includegraphics[width = 1\columnwidth]{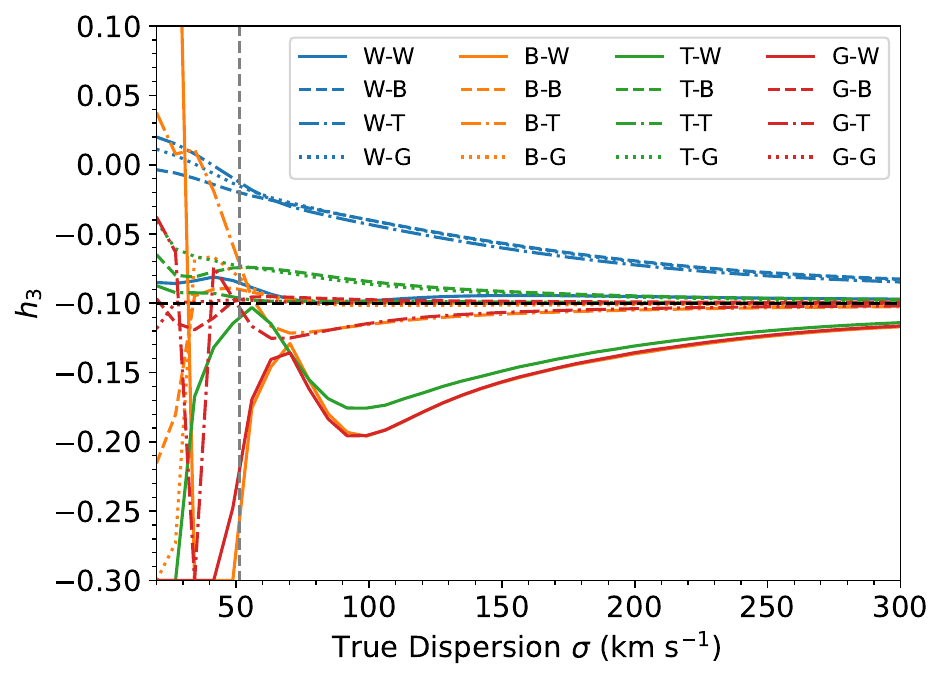}}
	\caption[]{Recovered \textsc{pPXF} $h_3$ for an underlying stellar LOSVD with $h_3 = -0.1$ and $h_4 = 0.1$. The MUSE instrumental dispersion is the dashed gray line at $\sigma = 51.4$ km s$^{-1}$. LSFs involving the winged profile are systematically too large or too small by up to $\sim0.08$ depending on if it is the template or target spectrum convolved with it. The remaining choices recover $h_3$ close to the instrumental resolution with errors of around 0.03.}
	\label{fig:standard_h3}
\end{figure}

\begin{figure}
	\centering
	\subfloat{\includegraphics[width = 1\columnwidth]{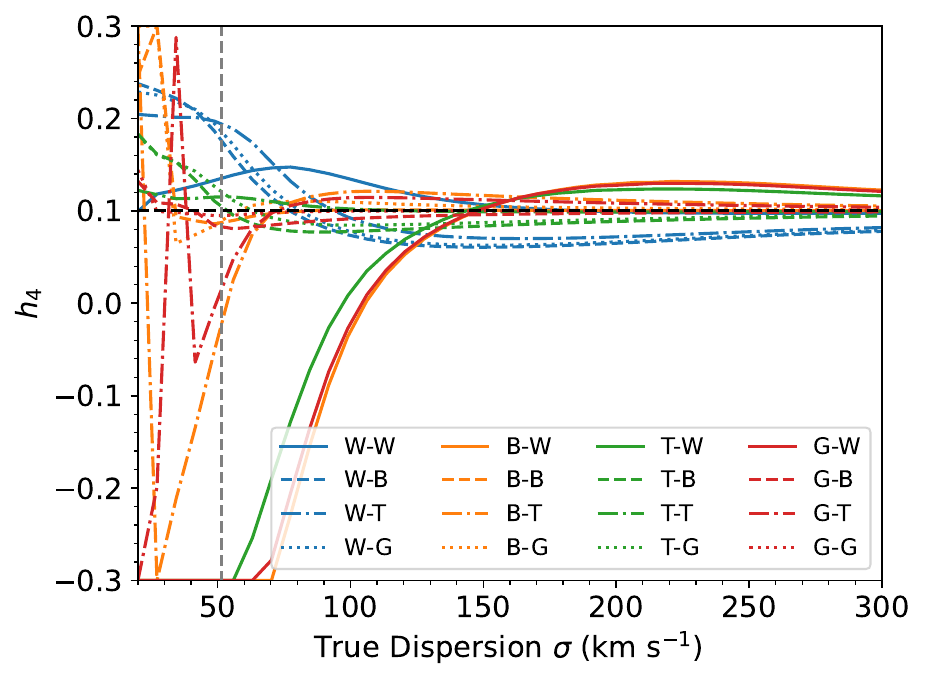}}
	\caption[]{Recovered \textsc{pPXF} $h_4$ for an underlying stellar LOSVD with $h_3 = -0.1$ and $h_4 = 0.1$. The MUSE instrumental dispersion is the dashed gray line at $\sigma = 51.4$ km s$^{-1}$. LSFs involving the winged profile are systematically too large or too small, with $h_4$ dropping to the smallest possible value of -0.3 when the template spectrum is convolved with it. The remaining choices recover $h_4$ close to the instrumental resolution with errors of around 0.03.}
	\label{fig:standard_h4}
\end{figure}

Here, I repeat the analysis from the previous part but assume the true stellar LOSVD is a GH series that has $h_3 = -0.1$ and $h_4 = 0.1$ with varying dispersion. I plot the recovered values of $h_3$ and $h_4$ as a function of the true stellar dispersion in \autoref{fig:standard_h3} and \autoref{fig:standard_h4}, respectively. The behavior here is noticeably different from the case where the underlying kinematics were a pure Gaussian, with a much larger range of recovered values at the large dispersion end. Here, the \simon{bias} in $h_3$ varies from about 0.02 at the large dispersion end to 0.15 at the instrumental dispersion. If the winged profiles are excluded then this reduces to 0.003 and 0.03, respectively. For $h_4$ the results are similar, with the total bias in the recovery ranging from 0.02 to 0.4, and 0.005 to 0.12 when winged profiles are excluded. It is worth emphasising that near the instrumental dispersion end it is generally not possible even in ideal conditions to accurately recover the GH moments as the LOSVD is undersampled. 

\begin{figure}
	\centering
	\subfloat{\includegraphics[width = 1\columnwidth]{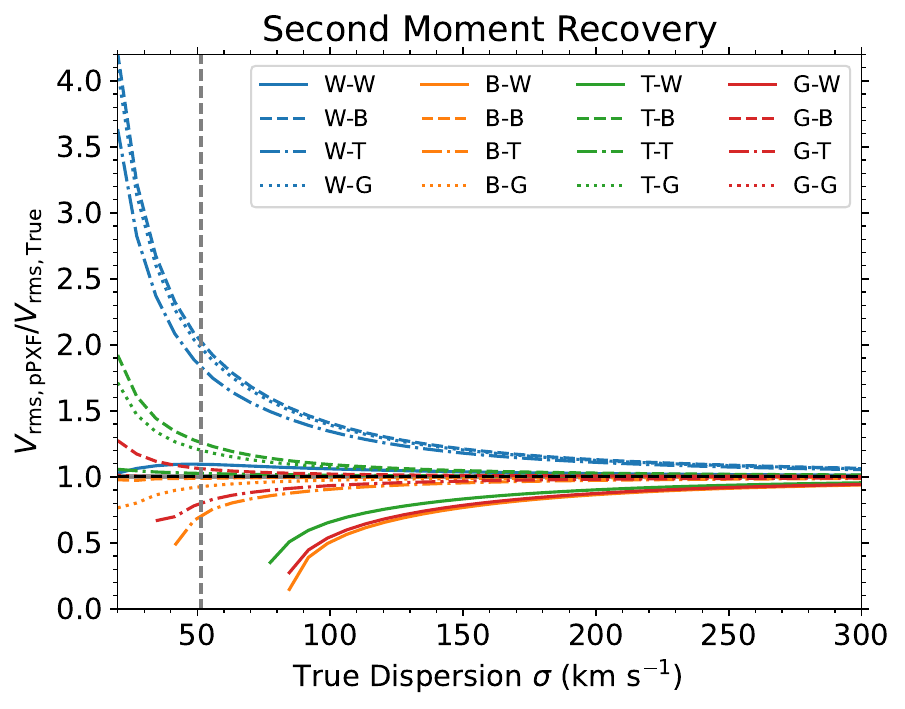}}\\
	\subfloat{\includegraphics[width = 1\columnwidth]{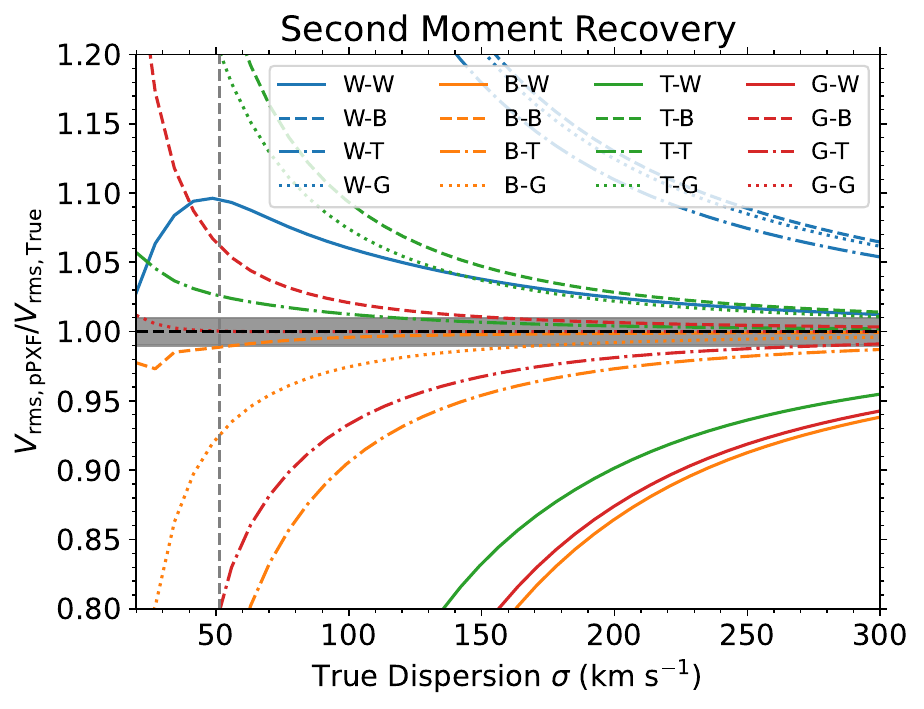}}
	\caption[]{Recovered \textsc{pPXF} $V_{\rm rms}$ for an underlying stellar LOSVD with $h_3 = -0.1$ and $h_4 = 0.1$. The MUSE instrumental dispersion is the dashed gray line at $\sigma = 51.4$ km s$^{-1}$. LSFs involving the winged profile are systematically too large or too small. Missing data points for the lower curves is due to the fact that $h_4$ becomes so negative that $V_{\rm rms}^2$ becomes negative. \simon{The top panel shows a zoomed out view of the data while the bottom panel shows a zoomed in view with a shaded gray region spanning one percent uncertainties. Several cases have sub-percent bias at the largest dispersions, though this is never the case for LSFs involving the winged profile.}}
	\label{fig:2mom}
\end{figure}

One unexpected result is that the recovered $h_4$, while initially larger than the true value for target LSFs with winged profiles, quickly decreases, ultimately giving the lowest values of $h_4$ as the dispersion increases. This contradicts the previous case in \autoref{fig:h4_recov_gauss} where these profiles were always the largest. The reason for this is that the convolution of a GH LOSVD with a winged GH LSF induces a number of higher order GH moments in the convolved spectrum. The best fit $h_4$ is difficult to predict as the fit LOSVD can expand either by increasing the dispersion, or by increasing $h_4$. A better quantity to look at to confirm that the behavior is still as expected is the second moment, given by
\begin{equation}
	V_{\rm rms} = \sqrt{\int_{-\infty}^\infty v^2\mathcal{L}_{\rm norm}(v) dv} = \sqrt{\overline{V}^2 + \sigma^2}
\end{equation}
with $\mathcal{L}_{\rm norm}$ the normalised LOSVD, $\overline{V}$ the mean of the normalised LOSVD and $\sigma$ the standard deviation. Note that when $h_4$ is non-zero these are not the same as the parameters $V$ and $\sigma$ is the GH series. Evaluating the second moment and plotting gives \autoref{fig:2mom}. There it is clear that the expected behavior is recovered, with the winged target LSFs returning the largest second moments. 


\subsection{LSF Toy Model}
\begin{figure}
	\centering
	\subfloat{\includegraphics[width = 1\columnwidth]{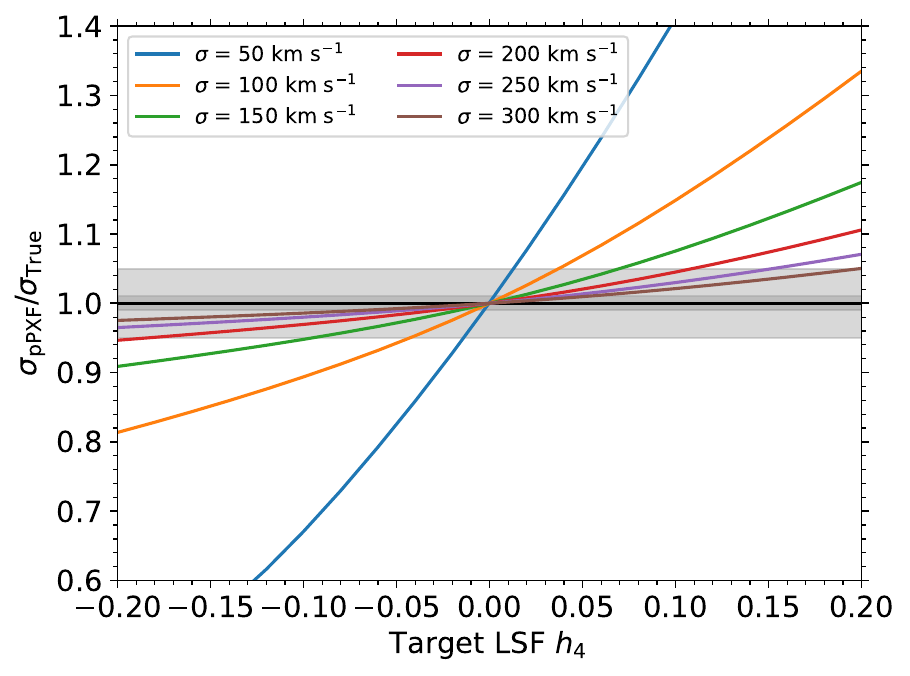}}\\
	\subfloat{\includegraphics[width = 1\columnwidth]{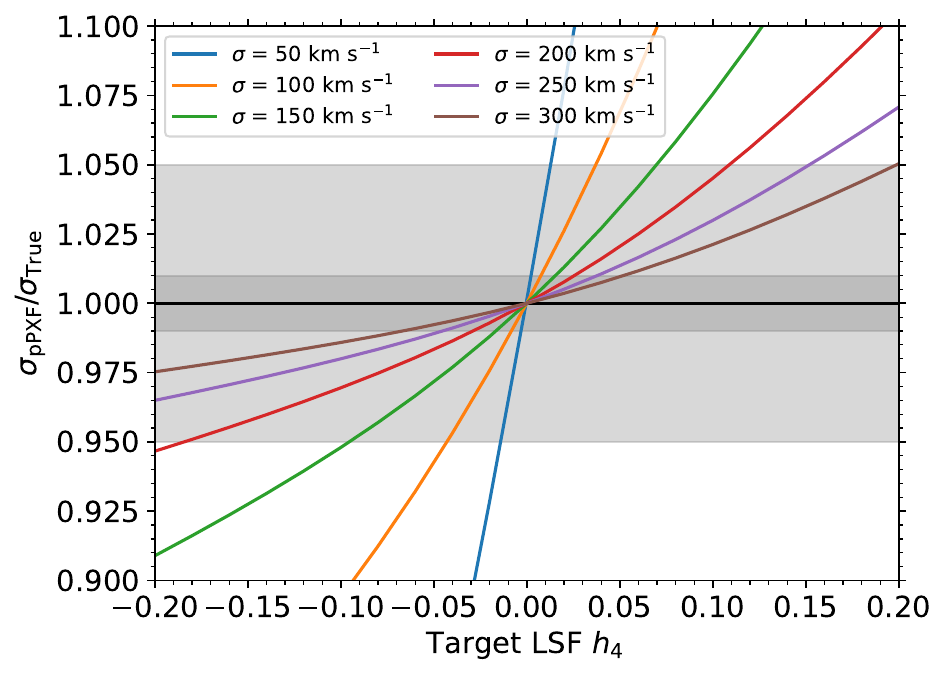}}
	\caption[]{Recovered \textsc{pPXF} dispersion for an underlying stellar LOSVD that is Gaussian. The LSF of the template spectrum is a pure Gaussian whereas the LSF of the target spectrum is a GH series with $h_4$ given by the $x$ axis. Shaded regions are drawn at the 99 percent and 95 percent accuracy levels. \simon{The bottom panel is a zoomed in version of the top panel.}}
	\label{fig:toy_lsf}
\end{figure}

So far I have run a number of tests studying how well-resolved non-Gaussian LSF profiles influence the recovery of the LOSVD. In general, however, detailed LSF profiles are not always available. One alternative is to measure the shape of the LSF from unresolved features like skylines or planetary nebulae \citep{isobe2023jwstlsf,shajib2025lsf}. Due to undersampling, it is not possible to extract a unique detailed LSF profile from this. However, the lowest order GH moments $h_3$ and $h_4$ can typically be measured as long as the features are well isolated. One example is \cite{vandesande2017higherorder}, which found median values of $h_3$ and $h_4$ less than 0.01 \simon{(see \autoref{sec:empirical_lsf})}. How does this translate into a \simon{bias} in the recovered dispersion? I investigate this by measuring the recovered dispersion where I fix the LSF of the templates to be a pure Gaussian and the LSF of the target spectrum to be Gaussian with variable $h_4$. I show the results of this for several choices of the dispersion in \autoref{fig:toy_lsf}. 

 \simon{One key difference between this and the work so far is that the dispersion and Gauss-Hermite moments measured in this way do not return the LSF on the linearly spaced domain $\lambda$ as in \autoref{eq:LSF_lambda} but rather as a function of the log rebinned spectrum. However, to lowest order the shape of the LSF is unchanged by log rebinning.} From this, we can see that if the SAMI survey is using well-calibrated template spectra with a Gaussian LSF, then the \simon{bias} on their dispersion due to the LSF is around 1 percent for $\sigma = 100$ km s$^{-1}$, suggesting that most of the SAMI survey would not be significantly impacted by the non-Gaussian shape of the LSF.

\begin{figure}
	\centering
	\subfloat{\includegraphics[width = 1\columnwidth]{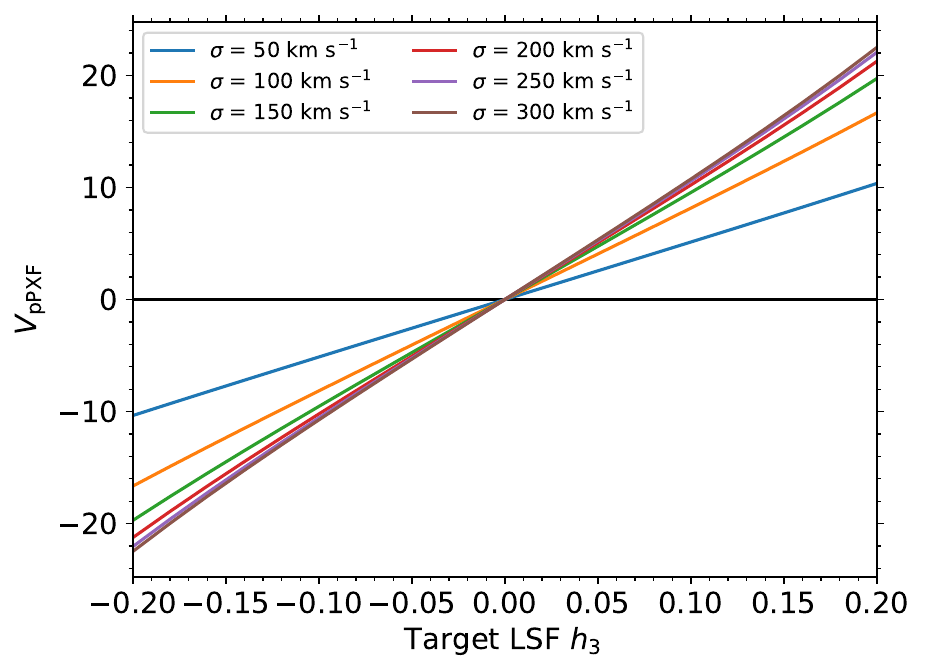}}
	\caption[]{Recovered \textsc{pPXF} velocity for an underlying stellar LOSVD that is Gaussian. The LSF of the template spectrum is a pure Gaussian whereas the LSF of the target spectrum is a GH series with $h_3$ given by the $x$ axis. \simon{We see a clear linear positive correlation between $h_3$ and $V$.}}
	\label{fig:toy_lsf_h3}
\end{figure}

I repeat this test but for $h_3$ for the velocity recovery and show the result in \autoref{fig:toy_lsf_h3}. As previously observed, it is the fits with large stellar dispersion that have the most deviant measured velocities. The dependence on $h_3$ is approximately linear. 

I also tested how the recovered velocity depends on $h_4$ and how the recovered dispersion depends on $h_3$, but found the impact to be negligible (less than 3 percent impact on the dispersion and less than 0.1 km s$^{-1}$ impact on the velocity). This makes intuitive sense as the velocity is primarily coupled to $h_3$ and the dispersion to $h_4$.

\section{Mitigation Strategies}\label{sec:mitig_strat}
\subsection{Special Spectral Libraries}
There are two clear methods for mitigating the impact of non-gaussian LSF profiles. A conceptually simple approach is to use the same instrument to obtain both the IFS cube and the template stars used for spectral fitting. Assuming that there is no spatial variations in the LSF, the template star will typically be taken to be the sum of all of the spaxels within a given aperture. In the case where there are spatial variations in the LSF \simon{(but see \autoref{sec:caveats})}, this would require making multiple pointings so that light from the star is in all spaxels of the IFS cube. Then the target spectrum in a spaxel would be fit with the template stars in the same spaxel. This approach perfectly matches the LSF of both the template and target spectra (up to time/configuration dependent changes such as gravitational flexure or impacts from the classical slit effect), implying that any observed broadening is due purely to the kinematics.  

This approach has the advantage of being conceptually simple and guaranteeing a treatment of the LSF that should be accurate to within the uncertainties of the spectral resolution. However, there is the disadvantage that it requires many observations to observe just a single template star, and in general many stars are needed in order to reproduce a given observed spectrum without introducing significant biases from template mismatch. 

\subsection{LSF Convolution Correction}
Another approach to minimising the uncertainty due to non-Gaussian LSF profiles is to perform detailed measurements of the LSF and incorporate this into the spectral fitting pipeline. This would ideally require obtaining high spectral resolution measurements of the LSF for both the target spectrum as well as for the template spectra used in the spectral fits. One then finds the function to convolve with the template LSF that reproduces the target LSF. This problem mathematically can be posed as solving
\begin{equation}\label{eq:conv_corr}
	\text{LSF}_{\rm Template}(\lambda)*G(\lambda) = \text{LSF}_{\rm Target}(\lambda)
\end{equation}
where $G$ is a function which, when convolved with the LSF of the template spectrum, returns the LSF of the target spectrum. I present a solution to this problem in \autoref{sec:solve_conv}.

\section{Solving the LSF Convolution Correction}
\label{sec:solve_conv}
\subsection{Wavelength Independent Solution}
In general, there is not a simple closed form solution to \autoref{eq:conv_corr} in the case that the LSF is wavelength dependent. However, In the wavelength independent situation, this is a well known special case of a Fredholm integral equation of the first kind. Using the Fourier convolution theorem, $G$ can be solved for exactly, giving
\begin{equation}\label{eq:wave_ind}
	G = \mathcal{F}^{-1}\left[ \frac{\mathcal{F}( \text{LSF}_{\rm Target})}{\mathcal{F}(\text{LSF}_{\rm Template})} \right]
\end{equation}
where $\mathcal{F}$ is the forward Fourier transform and $\mathcal{F}^{-1}$ is the inverse Fourier transform. For a pure Gaussian LSF this can be evaluated analytically giving \autoref{eq:gauss_conv}. However, when the LSF is parameterised as a Gauss-Hermite series there is no simple analytic solution to this equation and thus it must be approached numerically (see \autoref{sec:convolve_gh} for a special case). Despite this difficulty, this can still be solved with very high accuracy numerically. This is done by first using the analytic Fourier transform to evaluate $\mathcal{F}(\text{LSF}_{\rm Target})$ and $\mathcal{F}(\text{LSF}_{\rm Template})$ (see \cite{cappellari2017improving}). Note that while this equation provides a function in the same space as the spectrum, in practice it is more convenient to not take the inverse Fourier transform and perform the convolution directly in Fourier space using a fast fourier transform. 


\subsection{Wavelength Dependent Solution}
\label{sec:wave_dep}
In general there is not a simple closed form solution to \autoref{eq:conv_corr} in the case where the LSF is wavelength dependent. Several possibilities are available for a numerical implementation. The most direct option is to calculate the wavelength dependent convolution as a matrix multiplication with the spectrum. However, the convolution applied here will in most cases always be severely undersampled. This is because the scale of the convolution is $\sqrt{\sigma_{\rm Target}^2 - \sigma_{\rm Template}^2}$ which for MUSE and E-MILES varies from 0.6 \AA \ down to 0.05 \AA. This is a very small fraction of the 1.25 \AA \ spectral pixel size and so, even when oversampling, we expect a direct matrix multiplication implementation to suffer from this undersampling (see Fig. 2 of \cite{cappellari2017improving} for another example). For this reason, the approach must use the analytic GH Fourier transform based convolution. 

With this constraint in mind, another approach is possible. In the case that the LSF varies slowly as a function of the wavelength, an approximate solution can be obtained by segmenting the template spectrum into different chunks and applying the wavelength independent solution from the previous section at the mean wavelength value of the segment. These segments can then be added up to produce the wavelength dependent convolved spectrum. The key question is how to optimally segment the spectrum. Ideally one would like to smoothly segment the spectrum, allowing some overlap between neighboring segments so that the solution blends in the overlaping region. One natural choice for segmentation then is to use the Hanning window (or Hann function) \citep{blackman1958measurement}. The Hanning window is defined as
\begin{equation}
	w(x) = 
	\begin{cases}
		\frac{1}{L}\cos^2\left( \frac{\pi x}{L} \right) & |x| \leq L/2\\
		0 & |x| > L/2
	\end{cases}
\end{equation}
with $L$ the length of the window. The Hanning window has the nice property that windows with 50 percent overlap sum to a constant, making it easy to segment and sum up the spectrum. 

The detailed approach is then as follows: the spectrum is segmented into sections of length $L$ with each chunk given by
\begin{equation}
	T_k = w_k(x) \times T_{\rm Template}
\end{equation}
where $k$ denotes shifts by $L/2$. The wavelength independent problem is solved (\autoref{eq:wave_ind}) giving 
\begin{equation}
	\mathcal{F}(G_k) =  \frac{\mathcal{F}(\text{LSF}_{k,\rm Target})}{\mathcal{F}(\text{LSF}_{k,\rm Template})}
\end{equation}
where LSF$_{k}$ is the average LSF within the window $k$. Next, the segmented spectrum is convolved with this matching convolution, giving
\begin{align}
	T_{k,\rm Matched} &= T_k * G_k\\
	&= \mathcal{F}^{-1}\left( \mathcal{F}(T_k) \frac{\mathcal{F}(\text{LSF}_{k,\rm Target})}{\mathcal{F}(\text{LSF}_{k,\rm Template})} \right)\label{eq:t_matched}
\end{align}
While the Hanning function has the property that neighboring windows with 50 percent overlap sum to a constant, in general the GH series is not normalised and therefore a normalisation must be calculated. Additionally, the edges of the spectrum do not have neighboring Hanning windows and thus do not add to a constant. The normalisation can be calculated by determining the wavelength dependent convolved `spectrum' where the template spectrum is constant. Practically, this is the same as solving \autoref{eq:t_matched} with the following modification
\begin{equation}
	\text{Norm}_k = \mathcal{F}^{-1}\left( \mathcal{F}(w_k) \frac{\mathcal{F}(\text{LSF}_{k,\rm Target})}{\mathcal{F}(\text{LSF}_{k,\rm Template})} \right)
\end{equation} 
Thus, the wavelength dependent convolution is given by
\begin{equation}
	\label{eq:matching_convolution}
	T_{\rm Matched} = \frac{\sum_k T_{k,\rm Matched} }{\sum_k \text{Norm}_k}
\end{equation}

The last remaining question is what the optimal choice of $L$ is. In general the smallest reliable value should be chosen as this better satisfies the approximation that the LSF varies smoothly. I tested a variety of choices and found that $L = 5$ spectral pixels is the lowest value that still allows for accurate fits to the kinematics. A Python implementation of this \simon{algorithm} along with example code is publicy available at \url{https://github.com/dsimon45/LSF_Matching}.

\subsection{Corrected Kinematics}
In \autoref{sec:recov_kin} I showed the recovered kinematics when the LSF of the template spectrum is matched to the LSF of the target spectrum by applying a wavelength dependent Gaussian convolution. In this section I show the recovered kinematics when the template LSF is matched to the target LSF using \autoref{eq:matching_convolution}. 

\begin{figure}
	\centering
	\subfloat{\includegraphics[width = 1\columnwidth]{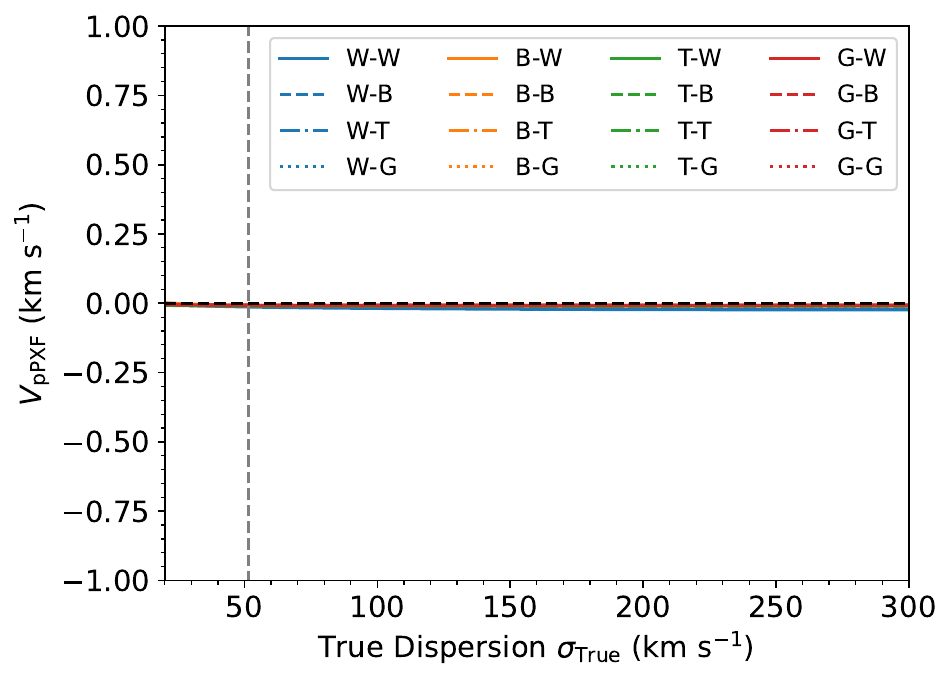}}
	\caption[]{Recovered \textsc{pPXF} velocity for an underlying LOSVD that is a pure Gaussian. The notation W-W means the target spectrum is convolved with the winged LSF profile and the template spectrum is also convolved with the winged LSF profile. The MUSE instrumental dispersion is the dashed gray line at $\sigma = 51.4$ km s$^{-1}$. The velocity is recovered everywhere to within $\pm0.05$ km s$^{-1}$.}
	\label{fig:improve_velocity}
\end{figure}

\begin{figure}
	\centering
	\subfloat{\includegraphics[width = 1\columnwidth]{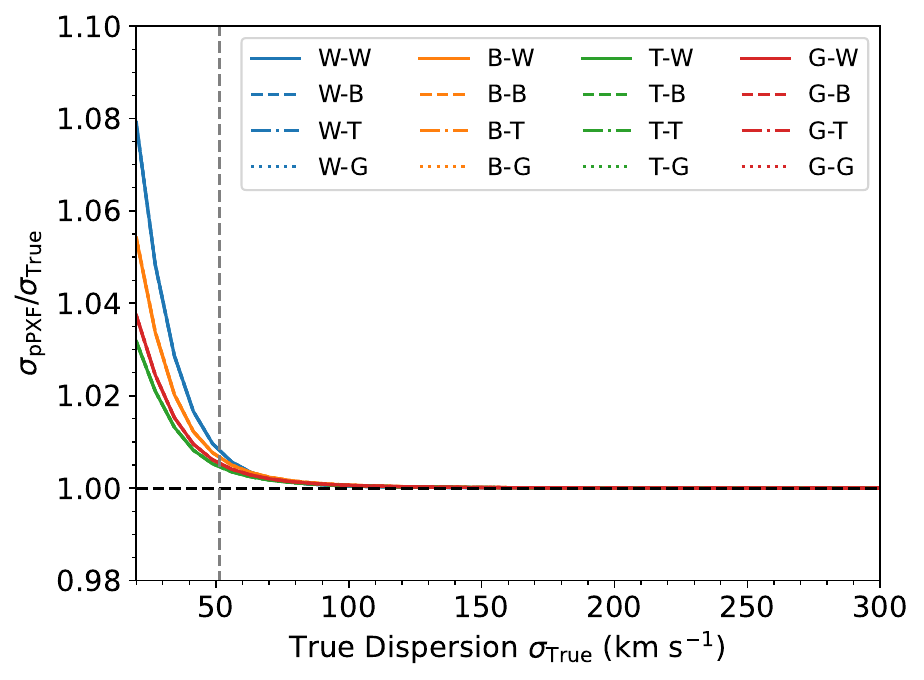}}
	\caption[]{Recovered \textsc{pPXF} $\sigma$ for an underlying LOSVD that is a pure Gaussian. The notation W-W means the target spectrum is convolved with the winged LSF profile and the template spectrum is also convolved with the winged LSF profile. The MUSE instrumental dispersion is the dashed gray line at $\sigma = 51.4$ km s$^{-1}$. The dispersion is recovered to better than one percent down to the instrumental dispersion }
	\label{fig:improve_dispersion}
\end{figure}

First, I show the results for the velocity and dispersion recovered when the underlying LOSVD is assumed to be Gaussian with $M=2$ in \autoref{fig:improve_velocity} and \autoref{fig:improve_dispersion}. The velocity is now recovered to within 0.05 km s$^{-1}$ for all values of the true dispersion. Likewise, the dispersion is recovered with less than 1 percent error down to the instrumental dispersion. Beneath the instrumental dispersion the error increases but is no more than 8 percent at half the instrumental dispersion. One key feature of these plots is that the 16 cases plotted collapse onto just four different curves, each one corresponding to the target galaxy having a winged, boxy, triangular, or Gaussian LSF. The reason for this is that the matching convolution effectively removes any memory of the template LSF, perfectly matching it to the target LSF. 

\begin{figure}
	\centering
	\subfloat{\includegraphics[width = 1\columnwidth]{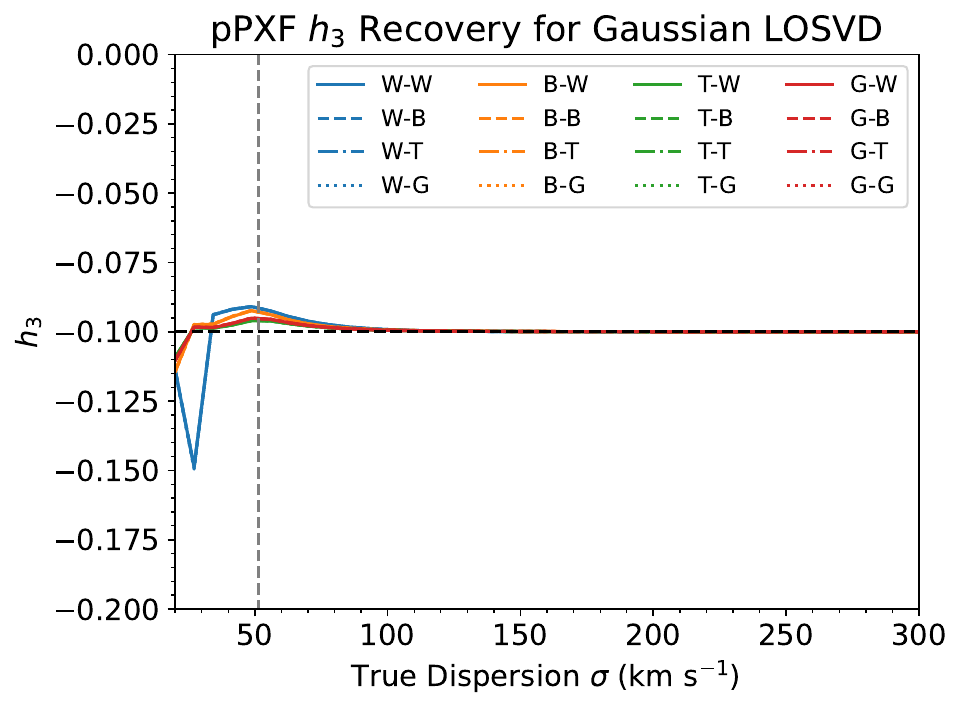}}
	\caption[]{Recovered \textsc{pPXF} $h_3$ for an underlying LOSVD that is a pure Gaussian. The notation W-W means the target spectrum is convolved with the winged LSF profile and the template spectrum is also convolved with the winged LSF profile. The MUSE instrumental dispersion is the dashed gray line at $\sigma = 51.4$ km s$^{-1}$.  $h_3$ is perfectly recovered down to twice the instrumental dispersion. At the instrumental dispersion it has a maximum error of around 0.02.}
	\label{fig:improve_h3}
\end{figure}

\begin{figure}
	\centering
	\subfloat{\includegraphics[width = 1\columnwidth]{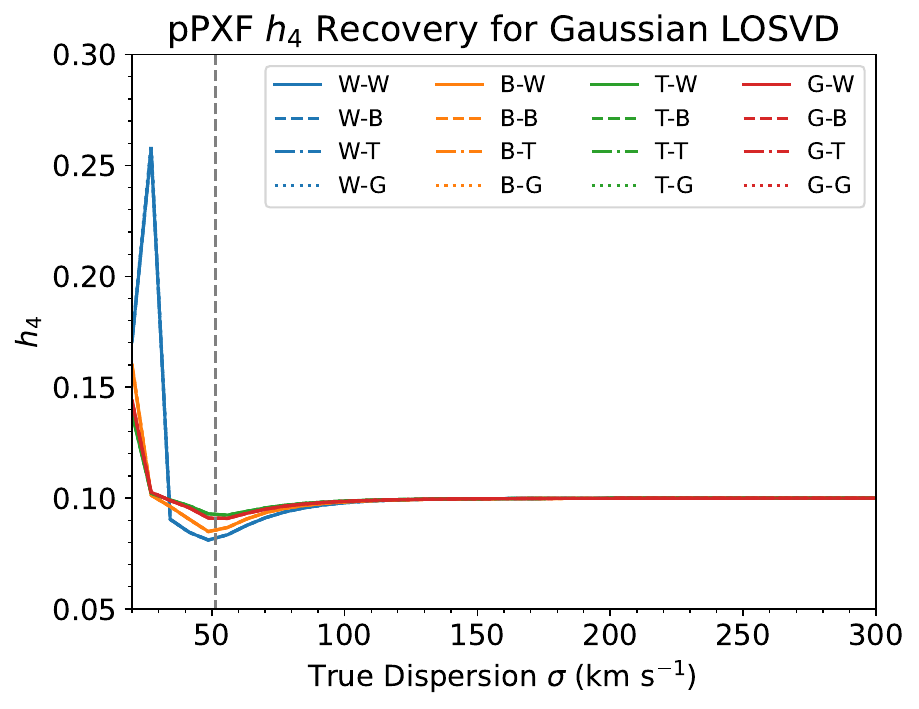}}
	\caption[]{Recovered \textsc{pPXF} $h_3$ and $h_4$ for an underlying LOSVD that is a pure Gaussian. The notation W-W means the target spectrum is convolved with the winged LSF profile and the template spectrum is also convolved with the winged LSF profile. The MUSE instrumental dispersion is the dashed gray line at $\sigma = 51.4$ km s$^{-1}$. $h_4$ is perfectly recovered down to twice the instrumental dispersion. At the instrumental dispersion it has a maximum error of around 0.02.}
	\label{fig:improve_h4}
\end{figure}

Next, I show the recovered $h_3$ and $h_4$ when the LOSVD is a GH series with $h_3 = -0.1$ and $h_4 = 0.1$ with $M=4$ in \autoref{fig:improve_h3} and \autoref{fig:improve_h4}. Again, we see that the 16 original cases collapse down to four. Both $h_3$ and $h_4$ are recovered with no offset down to twice the instrumental dispersion. Beneath this, they are recovered to within $\pm 0.02$ down to and a little beyond the instrumental dispersion. 0.02 is typical of the standard errors when measuring $h_3$ and $h_4$ (e.g. Fig. 7 of \cite{fu2025high}).

\begin{figure}
	\centering
	\subfloat{\includegraphics[width = 1\columnwidth]{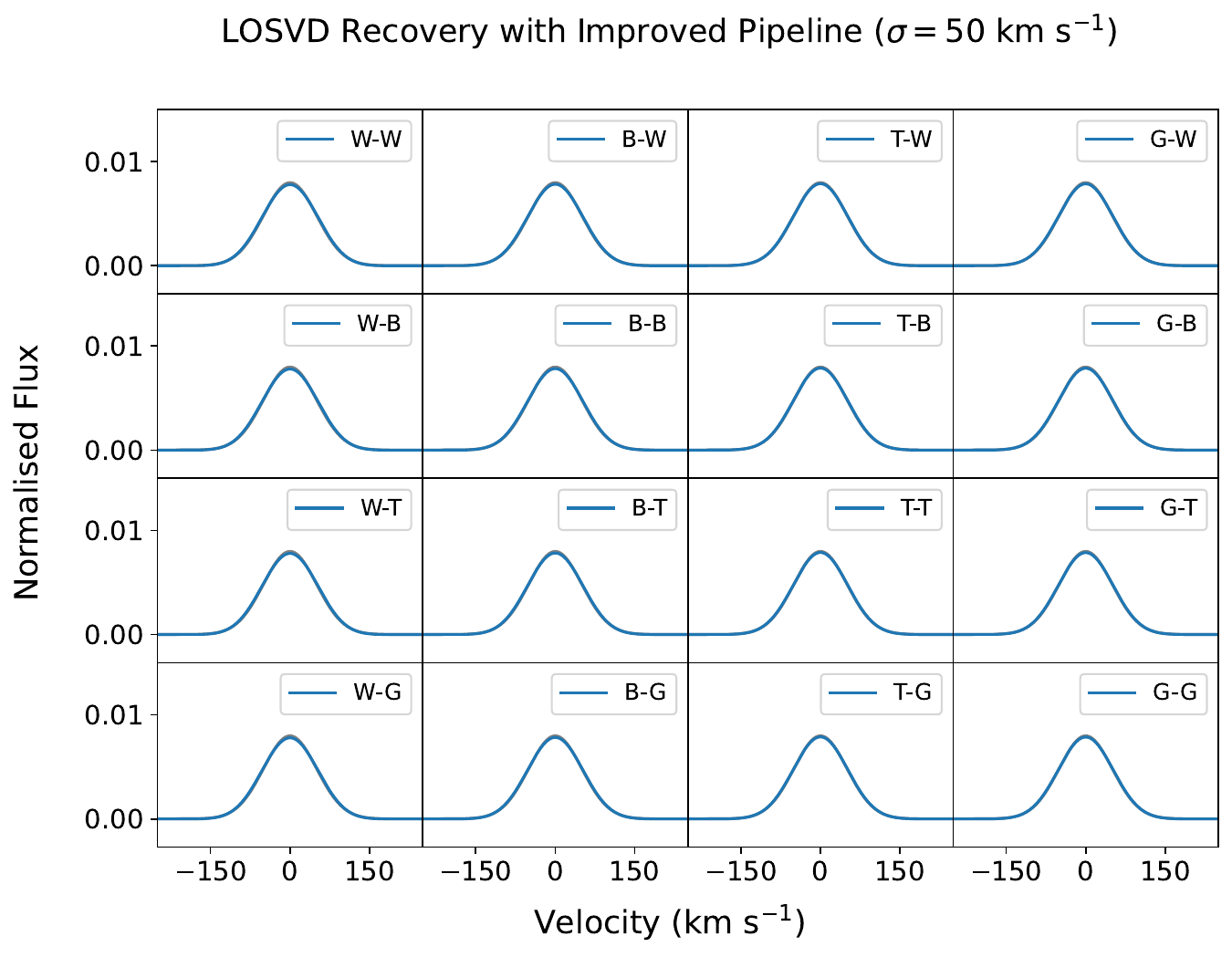}}
	\caption[]{Recovered \textsc{pPXF} LOSVD shapes with $M=4$ using the corrected LSF matching pipeline for an underlying stellar LOSVD that is a pure Gaussian with $\sigma = 200$ km s$^{-1}$. The true LOSVD is shown in gray. The notation W-W means the target spectrum is convolved with the winged LSF profile and the template spectrum is also convolved with the winged LSF profile. Now all of the recovered LOSVDs are nearly perfect even at the instrumental resolution. This stands in strong contrast to \autoref{fig:SS_losvd_recovery2} where the recovered LOSVDs have strong deviations.}
	\label{fig:SS_losvd_recovery3}
\end{figure}

\section{Discussion}\label{sec:discussion}
\subsection{Uncertainties in LSF Measurements}
Non-gaussian LSF shapes can significantly bias measurements of the kinematics of galaxies. Using the approach described in \autoref{sec:solve_conv}, it is possible to account for non-Gaussian shapes of the LSF, allowing accurate measurements of the LOSVD to be made. However, this assumes that the LSF profile is unchanged from the time it is measured to the time of the observation. This assumption was investigated for MaNGA in \cite{law2021mangalsf} where the authors obtained very well sampled LSF profiles using calibration lamps. They found that measurements of the width of the LSF varied by up to 10 percent when comparing measurements from skylines to measurements from calibration lamps. Additionally, standard stellar libraries such as those based on MILES find errors in their recovered FWHM of 2-4 percent (Fig. 4 of \cite{sanchez2006medium}, Fig. 5 of \cite{falcon2011updated}, Sec 4. of \cite{rock2015milesir}). This implies that even if the adjusted pipeline from \autoref{sec:solve_conv} is used, there may still be some bias in the recovered kinematics due to uncertainty in the detailed shape of the LSF. 

\begin{figure}
	\centering
	\subfloat{\includegraphics[width = 1\columnwidth]{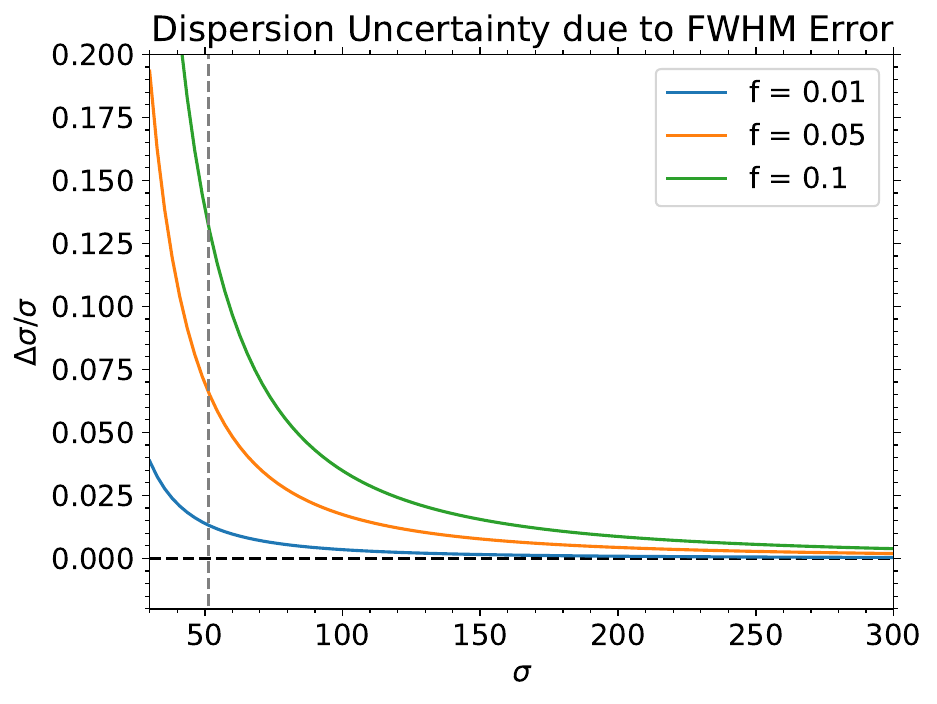}}
	\caption[]{Fractional uncertainty in the dispersion as a function of the true stellar dispersion for different values of the fractional error in the LSF width. The MUSE instrumental dispersion is the dashed gray line at $\sigma = 51.4$ km s$^{-1}$. Errors are typically on the order of a few percent.}
	\label{fig:dispersion_uncertainty}
\end{figure}

How exactly these errors propagate to measurements of the kinematics is not exactly clear. Reported uncertainties in the spectral resolution are typically averaged over the entire wavelength range. In reality, however, these uncertainties are comprised of different wavelength dependent uncertainties. If the uncertainty in the spectral resolution were normally distributed, then straightforward error propagation can be applied. The measured dispersion is given by
\begin{equation}
	\sigma^2_{*} = \sigma^2_{\rm Target} + \sigma^2_{\rm Target LSF} - \sigma^2_{\rm Template LSF} - \sigma^2_{\rm Matching LSF}
\end{equation}
The uncertainty on $\sigma_{*}$ is then given by
\begin{equation}
	\Delta \sigma_* = f \frac{\sqrt{\sigma_{\rm LSF \ Target}^4 + \sigma_{\rm LSF \ Template}^4}}{\sigma_*}
\end{equation}
with $f$ the fractional error on the LSF width. A plot of this function using the MUSE and E-MILES values of $\sigma_{\rm LSF \ Target}$ and $\sigma_{\rm LSF \ Template}$ is given for different values of $f$ in \autoref{fig:dispersion_uncertainty}. The typical uncertainties are of order a few percent at the instrumental dispersion, before asymptotically approaching zero. This suggests that, in most cases where there is a mismatch between the template and target LSF type, the uncertainty will be dominated by the shape of the LSF. However, the case where both template and target LSF have the same morphology (\autoref{fig:standard_dispersion_ww}) has comparable errors to this, suggesting that in this case the uncertainty in the \simon{FWHM} of the LSF becomes important and potentially dominant.

\subsection{Impact on Dynamical Models}
Recently there has been increased interest in measuring stellar velocity dispersions to very high accuracy \citep{knabel2025disperision}. Such efforts typically parameterise the line spread function as a wavelength dependent Gaussian for both the template and target spectra \citep{shajib2025lsf}. I have shown that in the extreme case even at dispersions of 300 km s$^{-1}$, there is up to a 7 percent \simon{bias} in the recovered dispersion if detailed LSF profiles are not taken into account. At 150 km s$^{-1}$ this is a 26 percent \simon{bias}. Accounting for detailed non-Gaussian LSF shapes allows for the dispersion to be recovered within these dispersion ranges to well within 1 percent accuracy. 

The downstream impact of this on dynamical models can be determined in a straight forward way. From the Jeans equation one can derive that the mass of the system depends on the second moment like $M \propto V_{\rm rms}^2$. From this, we see that a 7-26 percent uncertainty in the second moment corresponds to a 15-59 percent uncertainty in the recovered mass. This is a significant uncertainty, warranting a detailed study of the full shape of the LSF for both template and target spectra.
A natural question is whether or not such large \simon{biases} in the second moment have been observed in real galaxies. One relevant case is M87, which has been observed with NIFS \citep{gebhardt2011black}, OASIS, MUSE \citep{simon2023m87}, Keck KCWI \citep{liepold2023m87}, and JWST NIRSPEC \cite{al-amri2025jwstm87}. The measured velocity dispersion from each of these observations were compared in Figure 8 of \cite{al-amri2025jwstm87}. They find a variety of offsets between different observations. The JWST and Keck observations return the lowest dispersion and have little noticeable offset, while OASIS has the largest set of dispersions tending to be 25 percent higher than the JWST kinematics. The MUSE and NIFS observations are around 10 and 20 percent larger than the JWST dispersion, respectively. Of these, only \cite{liepold2023m87} measures the LSF, which they find to be boxy. \autoref{fig:standard_dispersion} shows that deviations for a boxy target LSF can reach 10 percent when the template spectra have broad wings. They fit their spectra with the MILES stellar library \cite{sanchez2006medium,falcon2011updated}. Measuring $h_3$ and $h_4$ for the MILES library would be an interesting test to determine to what extent, if any, non-Gaussian LSF profiles impact the recovery of the kinematics for M87. This, however, is beyond the scope of this work.

Additionally, it is important to emphasise that a variety of factors can contribute to the dispersion offset between two observations, including LSF variations but also template mismatch, degeneracies in the kinematic fits, sensitivity to additive or multiplicative polynomials, etc. \simon{In some cases the offset can even be physical, for instance if there are varying stellar populations tracing different kinematics.} LSF variations clearly have the potential to make up a substantial part of this uncertainty, but determining how much of that budget they occupy will require detailed measurements of the LSF for both template and target spectra.

One of the other ways that non-Gaussian LSFs can influence dynamical models is through their impact on the recovery of the higher order GH moments. As shown in \autoref{fig:h4_recov_gauss} and \autoref{fig:standard_h4}, fits to a target spectrum with non-Gaussian LSFs are forced to account for this, resulting in \simon{biases} in $h_4$ by up to 0.1. Uncertainties this large have been observed in real galaxies. One special case is NGC 1277 \citep{vandenbosch2012overmassive} for which an initial black hole mass measurement revealed a 1.7$\times 10^{10} M_\odot$ supermassive black hole mass. This was primarily due to the recovery of $h_4$ which peaked at a value of 0.2. A later approach with moments equivalent to a Jeans model found that missing the fits to $h_4$ by twice their uncertainty allows a much less massive black hole with $M_{\rm bh} = 5\times10^9 M_{\odot}$ to be recovered. \citep{emsellem2013overmassive}. This value was later confirmed with Schwarzschild modelling of NIFS data \citep{walsh2016ngc1277}, where the authors found that their measured $h_4$ was now half of the original values from \cite{vandenbosch2012overmassive}. Interestingly, this variation in $h_4$ of 0.1 is the max uncertainty introduced by LSF variations. At present it is not possible to determine how much of this (if any) was due to non-gaussian LSF shapes given that the detailed shape of the LSF for the spectrograph and templates are unknown. As this would be the most extreme case among the tests performed here, it suggests that other factors, such as template mismatch, also played an important role in the initial measurement of a large $h_4$. Detailed measurements of the LSF in the future for both stellar libraries and target spectra will give a better idea of how much of the observed scatter between different measurements of a galaxy's kinematics with different instruments is due to the LSF.

\subsection{\simon{Applications Beyond Stellar Kinematics}}
\simon{This work has solely focused on the impact that non-Gaussian line spread functions have on the recovery of stellar kinematics for MUSE like data. However, while the tests here are only performed in the context of stellar kinematic recovery, it is likely that the results would extend to a wider range of problems. The results should generalise well to different instruments, with the caveat that the key scale is the instrumental resolution, and this will vary by instrument. Additionally, emission line fitting will be subject to the same uncertainties of the LSF. The key difference is that the impact of the LSF is localised to the region of the emission line rather than coming from the entire wavelength range. Given that the tests here assume the LSF shape does not change as a function of wavelength (though the FWHM does vary), it is likely that the results here are directly applicable without modification to emission lines. }

\subsection{Caveats}
\label{sec:caveats}
As with any study, it is important to state and understand the limitations of the tests performed. I have assumed idealised conditions where the only uncertainty in the kinematic recovery comes from the shape of the LSF. In real observations, there will be other sources of uncertainty, such as atmospheric effects, template mismatch, contributions to the spectrum from non-stellar sources, \simon{finite signal to noise ratio,} etc. These have the potential to constructively or destructively combine with these results, further modifying the recovered kinematics. \simon{Though it is worth mentioning that, among these, low signal to noise ratio should not strongly impact the results here as the impact of noise is usually to randomly scatter the recovered parameters in the LOSVD, rather than to bias them \cite[Fig. 4]{cappellari2004parametric}. }

These tests have assumed mock observations that have a single non-Gaussian LSF shape that varies only in width as a function of the wavelength. Figure\simon{s 8 and} 9 of \cite{kakkad2022lsf} measure the GH moments of the LSF in a single spaxel for SINFONI and found that the GH moments do vary as a function of the wavelength. While these variations are not accounted for in the presented tests, they can be taken account of with the correction pipeline described in \autoref{sec:wave_dep}. The key assumption for that pipeline to work is just that the LSF varies smoothly across the spectrum. 

\begin{figure}
	\centering
	\subfloat{\includegraphics[width = 1\columnwidth]{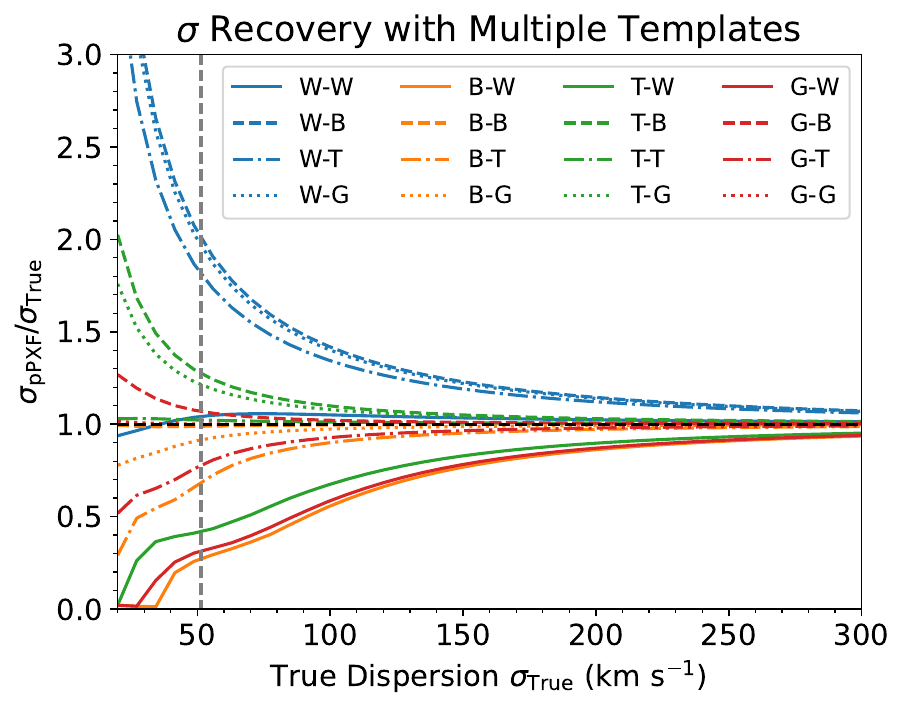}}
	\caption[]{Recovered \textsc{pPXF} $\sigma$ with $M=2$ for an underlying LOSVD that is a pure Gaussian. The full MARCS library with a Salpeter IMF is fit. These results are largely unchanged from \autoref{fig:standard_dispersion}, with some small variations of a few percent on the low dispersion side.}
	\label{fig:multitemp_dispersion}
\end{figure}

Additionally, in all of the tests shown I have assumed that the true spectrum is fit by a single template spectrum with the same SSP parameters as the target. This removes the freedom typically present in real analyses of IFS data for the template spectrum to consist of a linear combination that could potentially counteract the impact of the varying LSF. To test the impact this has on the recovered kinematics, I have performed the same test as in \autoref{sec:recov_kin} but fitting a linear combination of template spectra given by the entire MARCS library with a Salpeter IMF. The results for the recovered dispersion are shown in \autoref{fig:multitemp_dispersion}. The impact is small compared with the scale of the offsets, with the high dispersion range changing by a fraction of a percent and the low dispersion range changing by at most a few percent. This shows that using a linear combination of templates is not sufficient to rectify the impact of a varying LSF.

One final key detail of interest for future studies that has been ignored is the spatial variation of the LSF on a spaxel by spaxel level. Spatial variations of the LSF have been observed \cite[Fig. 7]{kakkad2022lsf}. The tests here fit a single template spectrum to a single target spectrum but for real galaxies the detailed systematic impact will depend on how well the LSF can be modelled across the detector. A fully realistic test would require having detailed LSF profiles across the detector for a real integral field spectrograph, as well as estimates of the error on the shape of the LSF. While this is important for highly accurate recoveries of the kinematics of low dispersion systems, this is beyond the scope of this paper.

\section{Conclusion}\label{sec:conclusion}
I have studied the impact of three non-gaussian LSF shapes on the recovery of the kinematics of a mock spectrum in an idealised setup. I summarise the primary results as follows:
\begin{itemize}
	\item The standard data reduction pipeline that ignores detailed LSF profiles has significant \simon{biases} even at \simon{six times the instrumental dispersion ($\sigma_{\rm inst} =$ 51 km s$^{-1}$)}, with \simon{biases} up to 7 percent at 300 km s$^{-1}$ and 26 percent at 150 km s$^{-1}$.
	\item \simon{Biases} in the recovered Gauss-Hermite moments are up to $\pm 0.1$ even when well resolved.
	\item A pipeline incorporating detailed LSF profiles can recover the dispersion with an accuracy of 1 percent down to the maximum instrumental dispersion.
\end{itemize}
Detailed LSF profiles have not previously been published for most integral field spectrographs or for most stellar template libraries. The systematic impact of not accounting for detailed LSF shapes depends significantly on the details of the LSF shape. Such measurements must be made in order to guarantee accurate stellar kinematic measurements for all values of the underlying dispersion.

A publicly available example code featuring a python implementation of the matching convolution described in \autoref{sec:solve_conv} is available at \url{https://github.com/dsimon45/LSF_Matching}.

\vspace{0.3cm}
\section*{Acknowledgments}
I thank Michele Cappellari for helpful discussions, and Niranjan Thatte for insightful discussions and for drawing my attention to the importance of LSF variations. I also thank Claudia Maraston for providing the MARCS SSPs. \simon{I would also like to thank the anonymous referees for comments that improved the clarity of this paper.} This research made use of \textsc{Jupyter}
\citep{PER-GRA:2007,Kluyver:2016aa}, \textsc{NumPy} \citep{harris2020array}, \textsc{SciPy} \cite{virtanen2020SciPy}, \textsc{Matplotlib}
\citep{hunter2007matplotlib}, \textsc{pPXF} \citep{cappellari2004parametric,cappellari2017improving,cappellari2022full}, and \textsc{WebPlotDigitizer} \citep{WebPlotDigitizer}. During the preparation of this manuscript, I used Claude (Anthropic) for proofreading and editing to improve language clarity and readability, as well as for code assistance and debugging. All AI-generated suggestions were critically reviewed, verified for accuracy, and adapted by the author, who takes full responsibility for the final content.


\bibliographystyle{mnras}
\bibliography{References}

\appendix

\section{Uncertainty in Wavelength Dependent Convolution}\label{sec:uncertain_conv}
\autoref{fig:standard_dispersion_ww} shows that even in the most simple case where the LSF for both the template and target spectrum is Gaussian, the stellar velocity dispersion can not be perfectly recovered beneath the instrumental dispersion (though the error is only 1-2 percent). The reason for this has to do with the numerical implementation of the wavelength dependent matching convolution.  

\begin{figure}
	\centering
	\subfloat{\includegraphics[width = .5\columnwidth]{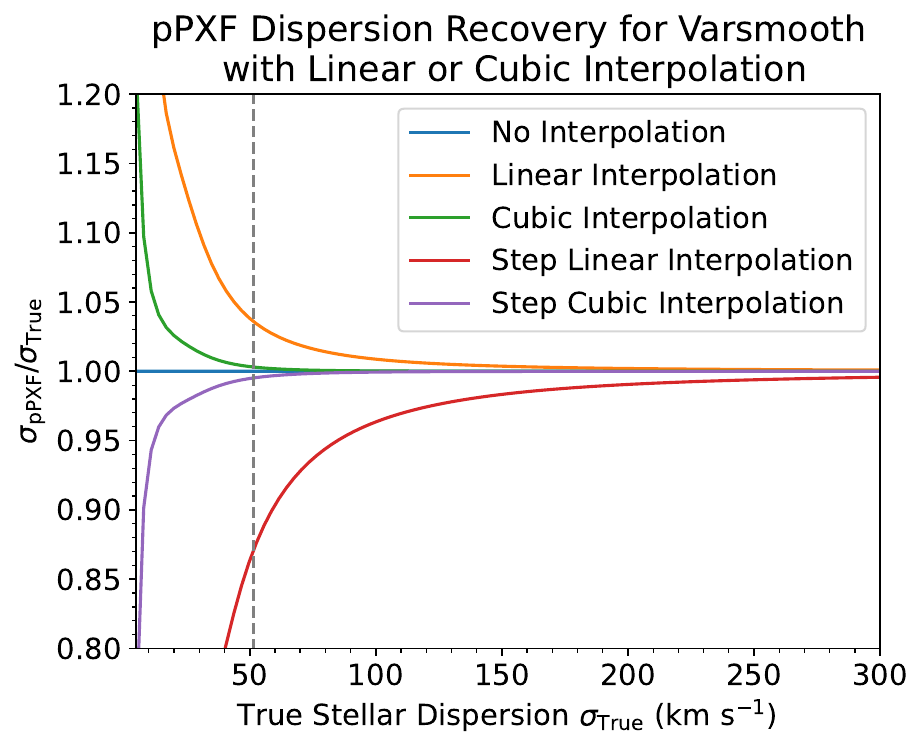}}
	\caption[]{Recovered \textsc{pPXF} $\sigma$ for an underlying LOSVD that is a pure Gaussian. The LSF of the template and target spectrum are both taken to be Gaussian with constant FWHM equal to 2.61 \AA\ for the target and 2.51 \AA \ for the template. For the curves labeled 'step' the template FWHM is 2 \AA \ and 2.5 \AA \ on the left and right hand halves of the spectrum, respectively. In the case that the LSF of the template is matched to the target by directly using the analytic Fourier transform, the dispersion can be perfectly recovered for any stellar velocity dispersion. When interpolation is used, an error is induced depending on the type of interpolation used.}
	\label{fig:interpolate_recovery}
\end{figure}

To show this, I consider a simplified scenario where the spectral resolution FWHM is constant for both the template and target spectrum with 2.51\AA \ for the template spectrum and 2.61 \AA \ for the target spectrum.. In this situation it is not necessary to do the wavelength dependent convolution (and interpolation) as the FWHM is wavelength independent, and the convolution can be evaluated in a straight forward way using the numerical implementation of the analytic Fourier transform. I show the recovered dispersion with \textsc{pPXF} for this in \autoref{fig:interpolate_recovery}. In this case the dispersion can be recovered perfectly all the way down to a tenth of the instrumental dispersion. To better understand the errors introduced by interpolation, I also repeat this measurement twice, but where I use the interpolation as described in algorithm 1 of \cite{cappellari2022full}. I do this using both linear and cubic interpolation and show the results in \autoref{fig:interpolate_recovery}. Both approaches perform well when the true stellar dispersion is large, but close to the instrumental dispersion it is clear that the approach with cubic interpolation out performs linear interpolation.

Linear and cubic interpolation have a mix of advantages and disadvantages. For noisy data, linear interpolation is likely to be more robust. Cubic interpolation provides a perfect fit to even noisy data, which can introduce spurious peaks and valleys on the interpolated grid. This effect is amplified if the interpolation is done on a non-uniformly spaced grid as uniform spacing limits how much the \simon{interpolated function can vary between neighboring points}. On smooth data, however, linear interpolation will systematically broaden convex and concave features. In that situation cubic interpolation ought to perform better.

In this test it is clear that cubic interpolation outperforms linear interpolation. The reason for this is likely that the template used is not noisy and the the spectral resolution varies smoothly, suggesting that cubic interpolation is justified in this case.

In practice, however, it is possible that the spectral resolution will not vary smoothly. One example of this is in the E-MILES stellar library, which stitches together stars from different observations in different spectral ranges. This leads to discrete jumps in the spectral resolution across the wavelength range of the library \cite[Fig. 8]{vazdekis2016emiles}. However, in this case it is still likely that cubic interpolation will outperform linear interpolation as this discontinuity will only introduce an uncertainty at one point in the spectrum. To test this, I repeat the above test but fix the template spectral resolution FWHM to be 2 \AA \ below 6975 \AA \ and 2.5 \AA \ above (the discontinuity is near the mid-point of the linearly spaced spectrum). The results of this are also shown in \autoref{fig:interpolate_recovery}. Even with the discontinuity in the spectral resolution, cubic interpolation out performs linear interpolation.

\section{Convolution of Gauss-Hermite Functions}\label{sec:convolve_gh}
Consider a non-gaussian LSF given by
\begin{equation}\label{eq:lsf_appendix}
	\text{LSF}(\lambda) = \frac{e^{\frac{-\Lambda^2}{2}}}{\sigma_\lambda \sqrt{2\pi}}\left[ 1 + h_3H_3(\Lambda) + h_4 H_4(\Lambda) \right]
\end{equation}
where $\Lambda = (\lambda - \lambda_0)/\sigma_\lambda$. The question is what the convolution of this function is with a Gaussian of the form
\begin{equation}
	\frac{1}{\sigma_*\sqrt{2\pi}}\exp\left[ \frac{-\Lambda^2}{2a^2} \right]
\end{equation}
where $a = \sigma_*/\sigma_\lambda$. I investigate each term separately. First, for the $h_3$ term we have
\begin{equation}
	\text{LSF}(\lambda) * G(\lambda) = 	\int_{-\infty}^{\infty}\frac{\exp\left[- \frac{(\Lambda-\Lambda^\prime)^2}{2a^2} - \frac{\Lambda^{\prime 2}}{2} \right]}{2\pi\sigma_*\sigma_\lambda} \left[ 1 + h_3H_3(\Lambda^\prime)\right] d\Lambda^\prime
\end{equation}
This convolution can be evaluated analytically using the properties of Gaussian integrals, and gives
\begin{equation}\label{eq:h3_conv}
		\text{LSF}(\lambda) * G(\lambda) = \frac{\exp\left(-\frac{\Lambda^2}{2(\sigma_x^2 + \sigma_\lambda^2)}\right) }{\sqrt{2\pi(\sigma_x^2 + \sigma_\lambda^2)}}\left[ 1 + \sqrt{3/2}a^4 h_3^\prime  H_1(\Lambda) + h_3^\prime H_3(\Lambda) \right]
\end{equation}
where 
\begin{equation}
	h_3^\prime = \frac{h_3 }{(1+a^2)^3}
\end{equation}
This has the usual behavior for the Gaussian term, though it also induces a $H_1(\Lambda)$ term and rescales $h_3$. A similar calculation can be made for $h_4$, giving
\begin{equation}
	\text{LSF}(\lambda) * G(\lambda) =\frac{\exp\left({-\frac{\Lambda^2}{2(\sigma_*^2 + \sigma_\lambda^2)}}\right)}{\sqrt{2\pi (\sigma_*^2+ \sigma_\lambda^2)}} \left[ 1 + \sqrt{\frac{3}{8}}h_4^\prime a^8 + \sqrt{3}a^4h_4^\prime H_2(\Lambda) + h_4^\prime H_4(\Lambda)\right]		
\end{equation}
where
\begin{equation}
	h_4^\prime = \frac{h_4}{(1+a^2)^4}
\end{equation}
Again, the Gaussian term is modified in the usual way, while also inducing an $H_0$ and $H_2$ term. As a check, both of expressions reduce to the appropriate terms in \autoref{eq:lsf_appendix} in the limit that $\sigma_* \to 0$. Higher order convolutions can be calculated (such as the case where both the LSF and $G$ contain some $h_4$), but this quickly becomes unwieldy.

\section{Increasing Velocity Recovery Model}\label{sec:inc_veloc}
\autoref{fig:standard_velocity} shows that the \simon{bias} in the recovered velocity increases as the true stellar dispersion increases. This is contrary to the case of the recovered dispersion where increasing the true stellar dispersion improves the accuracy of the recovery. In the second case this is because the convolution of the LSF with the velocity dispersion goes something like $\sqrt{\sigma_{\rm LSF}^2 + \sigma_{*}^2}$ so when $\sigma_{*}$ is large it will immediately dwarf the contribution from the LSF. It is less clear what the equivalent argument should be for the velocity.

To better understand this behavior, it is useful to have an analytic model. Consider the case of a Gaussian LOSVD and a LSF with some $h_3$. The convolution of these two functions together is given by \autoref{eq:h3_conv}. The measured velocity can be estimated by determining the velocity of this distribution. The general equation for this is
\begin{equation}\label{eq:velocity}
	\overline{V} \equiv \frac{1}{C}\int_{-\infty}^{\infty} v\mathcal{L}(v)dv
\end{equation}
where $\mathcal{L}$ is the LOSVD and $C$ is the normalisation given by 
\begin{equation}
	C = \int_{-\infty}^{\infty} \mathcal{L}(v) dv
\end{equation}
In our specific case $\overline{V}$ can be written as
\begin{equation}
	\overline{V} = \int_{-\infty}^{\infty} v\frac{e^{-y^2/2}}{\sqrt{2 \pi}\sigma} (1 + h_1 H_1(y) + h_3 H_3(y))dy
\end{equation}
with $y = (v-V)/\sigma$. This evaluates to the simple expression
\begin{equation}\label{eq:mean_vel}
	\overline{V} = \frac{(\sqrt{2}h_1 + \sqrt{3}h_3)\sigma}{C} + V
\end{equation}
All that remains is the evaluate the normalisation $C$. Parameterising this as a general Gauss-Hermite series gives 
\begin{equation}
	C = \int_{-\infty}^{\infty} \frac{e^{-y^2/2}}{\sqrt{2 \pi}\sigma}\left( \sum_{m=0}^{M}h_mH_m(y) \right)dy
\end{equation}
Taking $M=4$, this can be evaluated and gives
\begin{equation}
	C = h_0 + \frac{h_2}{\sqrt{2}} + \frac{1}{2}\sqrt{\frac{3}{2}}h_4
\end{equation}

\begin{figure}
	\centering
	\subfloat{\includegraphics[width = .5\columnwidth]{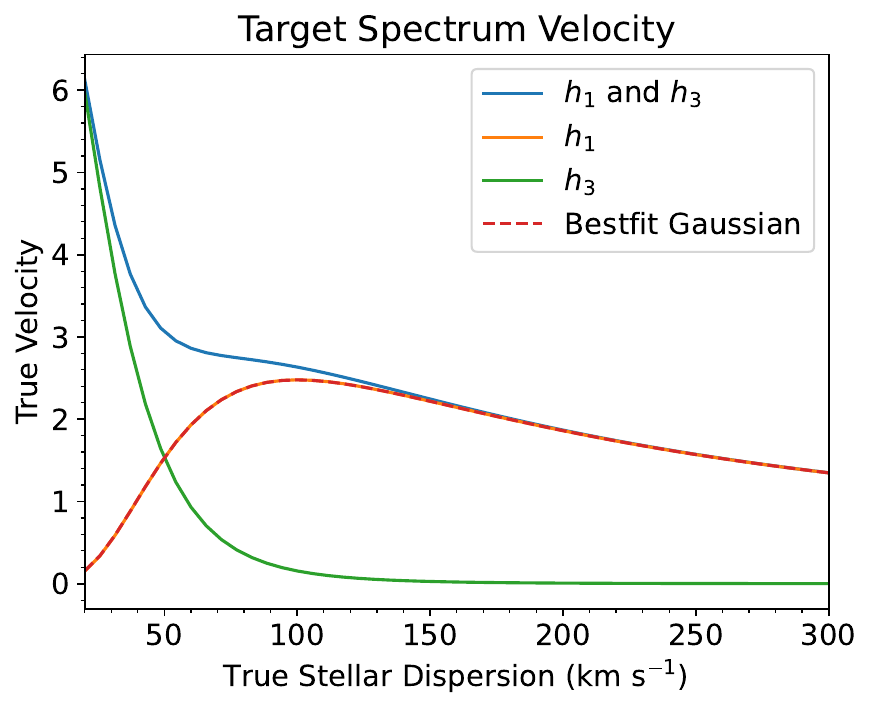}}
	\caption[]{The true velocity (or mean) of \autoref{eq:h3_conv} as a function of the stellar dispersion $\sigma_*$. The blue curve shows \autoref{eq:mean_vel} whereas the orange and green curves break this down into the contributions from just the $h_1$ term and the $h_3$ term. The $h_1$ term slowly increases before decreasing whereas the $h_3$ term uniformly decreases. Fitting a Gaussian to \autoref{eq:h3_conv} gives the dashed red curve.}
	\label{fig:h3_conv}
\end{figure}

Substituting in the values of $h_1$ and $h_3$ from \autoref{eq:h3_conv} and plotting as a function of $\sigma_*$ gives \autoref{fig:h3_conv}. There we see that the velocity does monotonically go to zero as the dispersion increases, but slowly. However, the behavior of each term in the velocity is quite different. The $h_3$ term starts out large and quickly goes to zero. The $h_1$ term, however, behaves exactly like the results in \autoref{fig:standard_velocity}, with the velocity starting low and then increasing before flattening off and decreasing again. 

This is the true velocity of the target spectrum in this simple analytic case. What \autoref{fig:standard_velocity} is showing is not the true velocity but the velocity of the best fit Gaussian. To understand how the best fit Gaussian velocity responds I fit a gaussian to \autoref{eq:h3_conv} as a function of $\sigma_*$ and plot the result in  \autoref{fig:standard_velocity}. The bestfit Gaussian velocity almost perfectly matches the contribution to the velocity from the $h_1$ term. The explanation for this is that the $h_1$ term is a lower order term in the series expansion so its contribution to the overall shape of the LOSVD is more central. This naturally has a higher impact on the centering of the bestfit Gaussian as compared to $h_3$.

In summary, the reason why the velocities recovered in \autoref{fig:standard_velocity} start low and increase is that the mean of the bestfit Gaussian approximately looks like $\sqrt{2}h_1\sigma$. As $\sigma_*$ increases $\sigma$ initially increases faster than $h_1$ term, causing the rise. However, eventually this changes as the $h_1$ decreases faster than $\sigma$, causing the velocity to eventually go to zero for very large values of $\sigma$.


\end{document}